\documentclass[traditabstract]{aa} 

\usepackage{graphicx}
\usepackage{txfonts}
\usepackage{longtable}
\begin{document}
   \title{The quenching of the star formation activity in cluster galaxies}
   \subtitle{}
  \author{A. Boselli\inst{1}\thanks{Visiting astronomer at CSIRO Astronomy and Space Science, Australia Telescope National Facility, PO Box 76, Epping, NSW 1710, Australia},         
	  Y. Roehlly\inst{1},
	  M. Fossati\inst{2,3},
	  V. Buat\inst{1},
	  S. Boissier\inst{1},
	  M. Boquien\inst{4},
	  D. Burgarella\inst{1},
	  L. Ciesla\inst{5},	  
	  G. Gavazzi\inst{6},
	  P. Serra\inst{7}       }

\institute{	
		Aix Marseille Universit\'e, CNRS, LAM (Laboratoire d'Astrophysique de Marseille), UMR 7326, F-13388, Marseille, France
             \email{alessandro.boselli@lam.fr, yannick.roehlly@lam.fr, veronique.buat@lam.fr, samuel.boissier@lam.fr, denis.burgarella@lam.fr}
        \and  
	        Universit{\"a}ts-Sternwarte M{\"u}nchen, Scheinerstrasse 1, D-81679 M{\"u}nchen, Germany
        \and
                Max-Planck-Institut f\"{u}r Extraterrestrische Physik, Giessenbachstrasse, 85748, Garching, Germany 
                \email{mfossati@mpe.mpg.de}
	\and
		Unidad de Astronomia, Universidad de Antofagasta
		\email{mederic.boquien@uantof.cl}
	\and
		Laboratoire AIM-Paris-Saclay, CEA/DSM/Irfu - CNRS - Universite Paris Diderot, CEA-Saclay, 91191, Gif-sur-Yvette, France
		\email{laure.ciesla@cea.fr}
	\and
		Universit\'a di Milano-Bicocca, piazza della scienza 3, 20100, Milano, Italy
		\email{giuseppe.gavazzi@mib.infn.it}
	\and
		CSIRO Astronomy and Space Science, Australia Telescope National Facility, PO Box 76, Epping, NSW 1710, Australia 
		\email{paolo.serra@csiro.au}
	}
               
\authorrunning{Boselli et al.}
\titlerunning{The quenching of the star formation activity of cluster galaxies}

   \date{}

 
  \abstract  
{We study the star formation quenching mechanism in cluster galaxies by fitting the spectral energy distribution of the \textit{Herschel} Reference Survey, 
a complete volume-limited $K$-band-selected sample of nearby galaxies including objects in different density regions, 
from the core of the Virgo cluster to the general field. The spectral energy distributions of the target galaxies are fitted using the CIGALE SED modelling code. 
The truncated activity of cluster galaxies is parametrised using a specific star formation history with two free parameters, the quenching age $QA$ and the
quenching factor $QF$. These two parameters are crucial for the identification of the quenching mechanism which acts on long timescales 
if starvation while rapid and efficient if ram pressure. To be sensitive to an abrupt and recent variation of the star formation activity, we combine 
in a new way twenty UV to far infrared photometric bands with three age-sensitive Balmer line absorption indices 
extracted from available medium-resolution ($R$ $\sim$ 1000) integrated spectroscopy and with H$\alpha$ narrow band imaging data.
The use of a truncated star formation history significantly increases the quality of the fit in HI-deficient galaxies of the sample, thus in those
objects whose atomic gas content has been removed during the interaction with the hostile cluster environment. The typical quenching age of the perturbed
late-type galaxies is $QA$ $\lesssim$ 300 Myr whenever the activity of star formation is reduced by 50\% $<$ $QF$ $\leq$ 80\% and
$QA$ $\lesssim$ 500 Myr for $QF$ $>$ 80\%, while that of the quiescent early-type objects is $QA$ $\simeq$ 1-3 Gyr. The fraction of late-type galaxies 
with a star formation activity reduced by $QF$ $>$ 80\% ~ and with an HI-deficiency parameter $HI-def$ $>$ 0.4 drops by a factor of $\sim$ 5 from the inner
half virial radius of the Virgo cluster ($R/R_{vir}$ $<$ 0.5), where the hot diffuse X-ray emitting gas of the cluster is located, to the outer regions ($R/R_{vir}$ $>$ 4).  
The efficient quenching of the star formation activity observed in Virgo suggests that the dominant stripping process is ram pressure. We discuss the
implication of this result in the cosmological context of galaxy evolution.
}
   {}
   {}
   {}
   {}
   {}

   \keywords{galaxies: clusters: general, galaxies: clusters: individual: Virgo, galaxies: evolution, galaxies: interactions, galaxies: ISM, galaxies: star formation
               }
   \maketitle
%

\section{Introduction}

Environment plays a major role in shaping galaxy evolution. Since the seminal work of Dressler (1980) it became
evident that galaxies in rich environments are systematically different than those located in the field. 
Dense environments are dominated by early-type galaxies, while low density regions by spirals and irregulars (e.g. Dressler
1980; Whitmore et al. 1993; Dressler et al. 1997). It has also been shown that massive local clusters are currently
accreting gas-rich, star forming systems (e.g. Tully \& Shaya 1984; Colless \& Dunn 1996) whose physical properties systematically 
change once they reach the densest regions. 
Indeed, it is now widely recognised that late-type galaxies located in rich clusters are generally HI-deficient 
(Haynes et al. 1984; Gavazzi 1987; Cayatte et al. 1990; Solanes et al. 2001; Gavazzi et
al. 2005, 2006a). There is a growing evidence indicating that they also lack of molecular gas 
(Fumagalli et al. 2009; Boselli et al. 2014b) and dust (Cortese et al. 2012a)
with respect to similar objects in the field. The angular resolution of the HI, CO, and far infrared images now available 
thanks to interferometric observations in the radio domain and to the superior quality of the instruments on board of 
\textit{Herschel} revealed that cluster galaxies have truncated gaseous 
and dust discs (Fumagalli et al. 2009, Boselli et al. 2014b, Cortese et al. 2012a, Davis et al. 2013), 
suggesting that the mechanism responsible for their stripping acts outside in. 
The activity of star formation of the late-type galaxies in clusters is also systematically 
reduced with respect to that of field objects (Kennicutt 1983; Gavazzi et al. 1998, 2002a, 2006b)
and it is limited to the inner disc 
(Koopmann \& Kenney 2004; Boselli \& Gavazzi 2006; Cortese et al. 2012b; Fossati et al. 2013). 
The correlation between the molecular gas content and the activity of star formation (e.g. Bigiel et al. 2008), generally called 
Schmidt law (Schmidt 1959; Kennicutt 1998a), can easily explain the observed truncation of the star forming discs of 
these perturbed objects.

Different mechanisms have been proposed in the literature to explain the transformation of galaxies in rich environments and the formation of the red
sequence (e.g. Boselli \& Gavazzi 2006, 2014). 
They include the gravitational interaction between galaxies (Merritt 1983) or that with the potential well of the cluster as a whole
(Byrd \& Valtonen 1990), or their combined effect generally called ``galaxy harassment'' (Moore et al. 1998). Other possible mechanisms
are those related to the interaction of the galaxy interstellar medium with the hot ($T$ $\sim$ 10$^7$-10$^8$ K) and dense 
($\rho_{ICM}$ $\sim$ 10$^{-3}$ cm$^{-3}$) diffuse gas trapped within the potential well of the clusters 
observable in X-rays (Sarazin et al. 1986). These include the ram pressure (Gunn \& Gott 1972) and the viscous stripping (Nulsen 1982)
exerted by the intracluster medium on galaxies moving at high velocity ($\sim$ 1000 km s$^{-1}$) within the cluster, or
the thermal heating of the galaxy ISM once in contact with the hot X-ray emitting gas of the cluster (Cowie \& Songaila 1977).
Since a large fraction of cluster galaxies is accreted via small groups, the perturbing mechanisms can start to shape galaxy evolution 
well before the galaxy is within the massive cluster (pre-processing, e.g. Dressler 2004). Finally, it is possible that under some circumstances 
the ICM only perturbs the hot gas in the galaxy halo unaffecting the cold component on the disc. The cold component, however, not replenished by
fresh infalling material, is later exhausted by the star formation activity of the galaxy itself (starvation; Larson et al. 1980).

The identification of the perturbing mechanism responsible for the quenching of the star formation activity of galaxies
in high density regions is becoming one of the major challanges of modern extragalactic astronomy. The identification of the dominant 
perturbing mechanism, which is expected to change with the mass of galaxies and with the properties of the overdensity region,
is crucial for cosmological simulations and semi analytical models of galaxy evolution. Different mechanisms may have different effects 
on the morphology and internal dynamics of the stellar component, making discs thicker and increasing the bulge-to-disc ratio (gravitational perturbations), 
while other are expected to decrease the surface brightness of the perturbed disc (starvation). At present, models and
simulations often overestimate the fraction of quiescent galaxies along the red sequence (Kang \& van den Bosch 2008; Font et al. 2008; Kimm et al. 2009;
Fontanot et al. 2009; Guo et al. 2011; Weinmann et al. 2011; Wang et al. 2012; Hirschmann et al. 2014; see however Henriques et al. 2015), suggesting that the physical prescriptions used
to reproduce the environmental quenching is still poorly understood. An accurate identification of the dominant mechanism 
is also crucial for the physical understanding of the stripping process, for the calibration of tuned hydrodynamic simulations, 
and for the charactrisation of the effects that the mechanism has on perturbed galaxies and on the stripped material.

Critical for constraining the perturbing mechanism is the identification with observations of its acting radius within 
high-density regions and of the quantification of the timescale necessary to significantly remove the gas and affect the 
star formation process (efficiency). The analysis of nearby and high redshift clusters still gives discordant results.
The detailed study of representative objects in nearby clusters such as Virgo and Coma based on dynamical 
modelling of their HI and CO gas kinematics (e.g. Vollmer et al. 2004), of the radial variation of their star formation properties
derived from integral field unit (IFU) optical spectroscopy (e.g. Crowl \& Kenney 2008), or by the comparison of multifrequency data with tuned 
chemo-spectrophotometric models of galaxy evolution (e.g. Boselli et al. 2006) indicate a recent and rapid truncation of their
star formation activity consistent with a ram pressure stripping scenario.
The analysis of large statistical samples of galaxies extracted from several surveys such as the SDSS, GAMA, or GALEX,
or targeted observations of nearby clusters and groups combined with the results of cosmological simulations or semi analytical 
models of galaxy evolution rather suggest a slow and long quenching process typical of starvation (McGee et al. 2009; Wolf et al. 2009; 
von der Linden et al. 2010; De Lucia et al. 2012; Wheeler et al. 2014; Taranu et al. 2014; Haines et al. 2015; Paccagnella et al. 2016). Other works suggest a bimodal evolution,
with an inefficient quenching at early epochs, when galaxies become satellites of more massive halos, and then a rapid quenching
when these still relative small systems are accreted in massive clusters (pre-processing; e.g. Wetzel et al. 2012, 2013; Muzzin et al. 2012;
Wijesinghe et al. 2012).

The identification of the perturbing mechanism requires an accurate reconstruction of the star formation history of galaxies in 
different environments. This is generally done using photometric and spectroscopic data to characterise the stellar emission 
in the UV to near-infrared spectral domain. The typical colours of galaxies in these bands are sensitive to their underlying stellar populations,
and become redder whenever the star formation activity decreses because perturbed by the surrounding environment.
The reconstruction of the star formation history of perturbed galaxies using these sets of data, however, is limited by the fact that other 
physical mechanisms such as dust attenuation and metallicity might affect the colours of galaxies in a similar way. To overcome this problem 
and limit any possible degeneracy between the effects of dust attenuation and aging of the stellar populations astronomers developed 
UV to far-infrared spectral energy distribution (SED) fitting codes. These codes, by measuring the total energy emitted by dust in the far-infrared,
are able to quantify in a self consistent way the dust attenuation on the stellar emission, and are thus now widely used to study the 
star formation history of different samples of galaxies (e.g. GRASIL: Silva et al. 1998; MAGPHYS: da Cunha et al. 2008).

In the last years a huge number of multifrequency data, from the UV to the radio centimetre, have been gathered for a complete
volume-limited sample of nearby galaxies, the \textit{Herschel} Reference Survey (HRS). This sample includes galaxies in 
different density regions, from the core of the Virgo cluster to the general field, and is thus perfectly suited for 
environmental studies (Boselli et al. 2010a). These data are perfectly suited to be fitted with CIGALE, a modelling code designed to 
reconstruct the star formation history of galaxies through an analysis of their UV to far infrared SED (Noll et al. 2009).
In this work we improve this SED fitting code to combine in an original manner twenty photometric bands  
imaging data with three age sensitive spectroscopic indices (Balmer absorption lines) derived from integrated spectroscopy 
(Boselli et al. 2013) and narrow band H$\alpha$ imaging data (Boselli et al. 2015) with the aim of characterising the quenching 
star formation episode. We already explored the use of the SED fitting analysis to reproduce the variations on the star formation history of
cluster galaxies using only broad band imaging photometry in Ciesla et al. (2016).
The present work is thus a further developement of this technique. To date the combination of photometric and spectroscopic data in the 
SED fitting analysis has been limited to the stellar emission (Pacifici et al. 2012, 2015; Newman et al. 2014; Thomas et al. 2016; Chevallard \& Charlot 2016;
Lopez Fernandez et al. 2016). Our work is the first
where this technique is extended to a much wider spectral domain, from the UV to the far-infrared, where the far-infrared emission is used through an energy balance to break 
any possible degeneracy due to dust attenuation in the observed stellar emission.
The paper is structured as follow:
we describe the sample in sect. 2, the multifrequency data in sect. 3, and the SED fitting technique and procedure in sect. 4. Section 4
also includes several tests done on mock samples to check the reliability of the output parameters. In sect. 5 we apply the
SED fitting code to a dozen of representative galaxies for which independent results are available in the literature, and in sect. 6 to
the whole HRS sample. The analysis is done in sect. 7 and the discussion in sect. 8. In Appendix A we discuss the effects of the adoption of different star
formation histories on the derived quenching parameters.  For a fair comparison with models and 
simulations we recall that the typical distance of the main body of the Virgo cluster is 17 Mpc 
(Gavazzi et al. 1999; Mei et al. 2007) and its total dynamical mass $M_{200}$ (1.4-4.2) $\times$ 10$^{14}$ M$_{\odot}$
(McLaughlin 1999; Urban et al. 2011; Nulsen \& Bohringer 1995; Schindler et al. 1999).

\section{The sample}

The sample analysed in this work has been extracted from the \textit{Herschel} Reference Survey (HRS;
Boselli et al. 2010a). The HRS is a volume-limited (15$\leq$ $D$ $\leq$25 Mpc), $K$-band-selected
(2MASS $K$ $\leq$ 12 mag for late-type galaxies, $K$ $\leq$ 8.7 mag for early-type
ellipticals and lenticulars) complete sample of 322 nearby galaxies. As selected, the sample spans 
a wide range in morphological type (E-S0-Spirals-BCDs-Im, see Table \ref{Tabtype}) and stellar mass (10$^8$ $\leq$ $M_{star}$
$\leq$ 10$^{11}$ M$_{\odot}$), making the HRS a statistically representative stellar mass selected 
sample of the nearby universe (Boselli et al. 2010a). The spectrophotometric data required for this analysis
are available for 168/260 late-type galaxies of the sample, while only for 22/62 early-type galaxies
(6/19 E, 15/36 S0, 1/4 S0a)\footnote{We exclude form the following analysis 
the radio galaxies M87 (Virgo A) and M84 because their far infrared emission is due to synchrotron (Baes et al. 2010; Boselli et al. 2010b, 2012).}.
The sample is ideally suited for environmental studies because it includes galaxies in a wide range
of densities, from the dense core of the Virgo cluster to small groups, binary systems and 
relatively isolated objects in the field. Strong observational evidence has been collected so far
indicating that the HRS galaxies in the Virgo cluster have been perturbed by their interaction with
the hostile environment (Cortese
et al. 2010, 2011, 2012a, 2012b, 2016; Boselli et al. 2014b, 2015). 
Although a few galaxies have a nuclear activity (Gavazzi et al. in prep.), 
the contribution of the AGN to the integrated spectrum is negligible (Boselli et al. 2013).
The HRS is also ideally defined because it includes a large number of well studied Virgo cluster
galaxies where the comparison of their kinematic and spectrophotometric properties have been crucial 
for the identification of the perturbing process and dating the epoch of the first interaction
with the hostile environment (Vollmer et al. 1999, 2000, 2004, 2005, 2006, 2008a, 2008b, 2009, 
2012; Vollmer 2003; Kenney et al. 2004; Boselli et al. 2005, 2006; Crowl \& Kenney 2008; 
Abramson et al. 2011; Kenney et al. 2014; Abramson \& Kenney 2014; Cortes et al. 2015).

\begin{table}
\caption{The morphological distibution of the HRS. }
\label{Tabtype}
{
\[
\begin{tabular}{ccc}
\hline
\noalign{\smallskip}
\hline
Type    &  HRS  & analysed galaxies$^a$     \\
\hline
E       & 19    & 6     \\
S0	& 36	& 15	\\
S0a	& 4	& 1	\\
Sa	& 27	& 13	\\
Sab	& 22	& 7	\\
Sb	& 52	& 39	\\
Sbc	& 33	& 19	\\
Sc	& 39	& 28	\\
Scd	& 30	& 26	\\
Sd	& 19	& 13	\\
Sm-Im-BCD& 39	& 23 	\\
\noalign{\smallskip}
\hline
\end{tabular}
\]
}
Notes: $a$ the galaxies analysed in this work have photometric data in all 
the following bands: $FUV$ and $NUV$ from GALEX, $gri$ from SDSS, $JHK$ from 2MASS, 
11 and 22 $\mu$m from WISE, 100 and 160 $\mu$m from PACS and 250, 350, 500 $\mu$m from SPIRE, 
as well as H$\alpha$ imaging data and Balmer absorption lines from integrated spectroscopy.
\end{table}

\section{The data}

The HRS is perfectly suited for an SED analysis because photometric and spectroscopic data
covering the whole electromagnetic spectrum are available for all galaxies. All photometric and spectroscopic 
data are integrated quantities and thus do not deserve any aperture correction, a probable source of 
systematic effects in perturbed galaxies given the outside-in truncation of the star forming disc.
The objects analysed in this work have been detected in all the following photometric bands: 
$FUV$ and $NUV$ (GALEX), $gri$ (SDSS), $JHK$ (2MASS), 
11 and 22 $\mu$m (WISE), 100 and 160 $\mu$m (PACS) and 250, 350, 500 $\mu$m (SPIRE), and
have H$\alpha$ imaging data and Balmer absorption line indices from integrated spectroscopy. Systematic effects due to 
different sensitivities in the various bands should thus be minimal.

\subsection{Photometry}

An accurate determination of the physical properties of galaxies through an SED modelling technique 
requires a full sampling of the electromagnetic spectrum from the UV bands (recent star formation
activity) to the near- (bulk of the stellar mass), mid- (dust attenuation), and far-infrared (dust mass) (e.g. Boselli
2011).
The HRS has been covered at all wavelengths for this purpose. Ultraviolet data in the $FUV$ ($\lambda$ 1539 \AA)
and $NUV$ ($\lambda$ 2316 \AA) have been collected thanks to two dedicated GALEX surveys (Boselli et al. 2011; Cortese
et al. 2012b). Optical data in the $gri$ bands have been extracted from the SDSS (Cortese et al. 2012b), while 
near-IR $JHK$ data from 2MASS (Skrutskie et al. 2006). Mid-IR data are available thanks to the WISE (11 and 22 $\mu$m) 
and \textit{Spitzer} (IRAC 8 $\mu$m) space missions (Ciesla et al. 2014). The sample has been also observed in the far-IR
with PACS (100-160 $\mu$m; Cortese et al. 2014) and SPIRE (250-350-500 $\mu$m; Ciesla et al. 2012) on 
\textit{Herschel}, while MIPS-\textit{Spitzer} data (at 24 and 70 $\mu$m) are available only for a 
fraction of the sample (Bendo et al. 2012).
Narrow band H$\alpha$+[NII] imaging data are available for the vast majority of the star forming galaxies of the sample 
(Boselli et al. 2015). 
The uncertainty in the different photometric bands slightly changes from galaxy to galaxy as shown in the original papers where the data are published. 
Typical uncertainties are given in Table \ref{Tabunc}.

\begin{table}
\caption{Uncertainties in the different photometric and spectroscopic bands. }
\label{Tabunc}
{
\[
\begin{tabular}{cccc}
\hline
\noalign{\smallskip}
\hline
Band		& Instrument	& Uncertainty	& ref     \\
\hline
$FUV$		& GALEX		& 15\%		& 1	\\
$NUV$		& GALEX		& 15\%		& 1	\\
$g$		& SDSS		& 15\%		& 1	\\
$r$		& SDSS		& 15\%		& 1	\\
$i$		& SDSS		& 15\%		& 1	\\
$J$		& 2MASS		& 15\%		& 2	\\
$H$		& 2MASS		& 15\%		& 2	\\
$K$		& 2MASS		& 15\%		& 2	\\
8$\mu$m$^a$ 	& IRAC/Spitzer	& 15\%		& 3	\\
11$\mu$m	& WISE		& 6\%		& 3	\\
22$\mu$m	& WISE		& 13\%		& 3	\\
24$\mu$m$^a$ 	& MIPS/Spitzer	& 4\%		& 4	\\
70$\mu$m$^a$ 	& MIPS/Spitzer	& 10\%		& 4	\\
100$\mu$m	& PACS/Herschel & 16\%		& 5	\\
160$\mu$m	& PACS/Herschel & 12\%		& 5	\\
250$\mu$m	& SPIRE/Herschel& 6\%		& 6	\\
350$\mu$m	& SPIRE/Herschel& 8\%		& 6	\\
500$\mu$m	& SPIRE/Herschel& 11\%		& 6	\\
H$\alpha$	& SPM		& 15\%		& 7	\\
H$\beta$ 	& OHP		& 15\%		& 8	\\
H$\gamma$ 	& OHP		& 15\%		& 8	\\
H$\delta$ 	& OHP		& 15\%		& 8	\\
\noalign{\smallskip}
\hline
\end{tabular}
\]
}
Notes: $a$ available only for a fraction of the analysed galaxies. \\
References: 1) Cortese et al. (2012b), 2) Jarrett et al. (2003), 3) Ciesla et al. (2014),
4) Bendo et al. (2012), 5) Cortese et al. (2014), 6) Ciesla et al. (2012), 7) Boselli et al. (2015), 8) Boselli et al. (2013).
\end{table}

\subsection{Spectroscopy}

Age-sensitive absorption Balmer line indices have been extracted from the integrated medium resolution ($R$ $\simeq$
1000) spectra of the star forming (Boselli et al. 2013) and quiescent (Gavazzi et al. 2004) galaxies of the HRS. 
These spectra, which cover the 3500-7000 \AA ~spectral domain, have been obtained by drifting the slit of the 
spectrograph over the disc of galaxies. These spectra are thus representative of the whole galaxies and can be combined 
in the SED fitting analysis with the photometric data without any aperture correction. To remove the contribution of emission lines, 
the reduced spectra are fitted using the GANDALF code (Sarzi et al. 2006; Falcon-Barroso et al. 2006) as described in Boselli et al.
(2015). This code has
been designed to simultaneously fit the emission and absorption lines to properly separate the relative contribution
of the stellar continuum from the nebular emission in the spectra of galaxies. Once removed the contribution of the
emission lines, the stellar continuum spectra are normalised\footnote{Because of the drifting observing technique, 
the spectra do not give absolute fluxes.} to match the total emission of galaxies using the 
monochromatic $g$-band flux density of Cortese et al. (2012b) by convolving the spectrum with the SDSS $g$-band filter
transmissivity. \\

\subsection{Corollary data}

The activity of star formation in galaxies is tightly connected to their total gas content (Boselli et al. 2001). The interaction of galaxies with the hostile cluster
environment removes the gaseous component inducing a quenching of the star formation activity. We thus compare the star formation
properties of the sample galaxies to their amount of gas.
Atomic and molecular gas data for the large majority of the sample have been
collected and homogenised in Boselli et al. (2014a).
Atomic gas masses are used to derive the
HI-deficiency parameter $HI-def$ defined as the difference in logarithmic scale between the expected and
the observed HI mass of a galaxy of given angular size and morphological type (Haynes \& Giovanelli 1984). The
HI-deficiency for all the HRS galaxies given in Boselli et al. (2014a) has been determined using the recent calibration of 
Boselli \& Gavazzi (2009).\\

HI data are also important for measuring the rotational velocity of the target galaxies, 
an essential parameter in the following analysis (see sect. 4).

\section{The SED modelling}

\subsection{The models}

The observed SED of the HRS galaxies are fitted using the CIGALE\footnote{http://cigale.lam.fr/} SED 
modelling code (Noll et al. 2009; Ciesla et al. 2016; Boquien et al. in prep.). Its application to the HRS
galaxies has been already described in detail in Ciesla et al. (2014, 2016). Here we just summarise the main properties of this code.
CIGALE produces synthetic UV to far infrared SED of galaxies using different stellar population synthesis models available in 
the literature to trace the stellar emission and different dust models or empirical templates to trace the dust emission.
The stellar and dust emissions are related by the dust attenuation which is estimated taking into account an energetic balance between the 
energy emitted by the different stellar populations absorbed by dust and re-emitted in the infrared. The model spectra are constructed
assuming different star formation histories, metallicities, and attenuation laws (see below). They are then compared to the multifrequency 
observations once convolved with the transmissivity profile of different photometric bands generally used in ground-based 
and space missions. The code identifies the best fitted model through a $\chi^2$ minimisation and makes
a probability distribution function (PDF) analysis to identify the likelihood-weighted mean value and standard deviation for
different physical parameters such as the stellar mass, the star formation rate, etc. This code has been successfully used to
reconstruct the star formation history of high-$z$ and local galaxies (Buat et al. 2014, Boquien et al. 2016, Lo Faro et al. in prep.), 
including those of the HRS (Boquien et al. 2012, 2013; Ciesla et al. 2014, 2016).\\

The novelty introduced in this work is the inclusion in the fitted variables of widely used age-sensitive spectral features extracted from 
medium resolution ($R$ $\sim$ 1000) spectra. For this purpose we use the high resolution version of the Bruzual \& Charlot population synthesis 
models (Bruzual \& Charlot 2003). Consistently with our previous works, the far infrared part of the spectrum is fitted with 
an updated version of the Draine \& Li (2007) physical models of dust emission. These models are characterised by different variables
tracing the dust properties: $q_{PAH}$, the fraction of the total dust mass in PAHs containing less than 10$^3$ C atoms, $U_{min}$, the intensity
of the diffuse interstellar radiation field, and $\gamma$, the fraction of dust heated by young stars within photodissociation regions (PDRs), as
extensively described in Draine \& Li (2007) and Draine et al. (2007).\\

\subsection{Parametrisation of the star formation history}

Several observational properties of local galaxies, including SED, colour and metallicity gradients, and different scaling relations, 
are well reproduced assuming that galaxies are coeval systems and evolved following 
a delayed star formation history (Sandage 1986; Boissier \& Prantzos 2000; Boselli et al. 2001; Gavazzi et al. 2002b). 
Different delayed star formation histories have been proposed in the literature.
In this work we adopt that derived by Buat et al. (2008) to reproduce the star formation history of the multizone chemo-spectrophotometric
models of galaxy evolution of Boissier \& Prantzos (2000). These models have been successful in reproducing the colour, metallicity and gas 
radial profiles of nearby galaxies such as those analysed in this work and their typical scaling relations at different redshift 
(e.g. Boissier \& Prantzos 2000, 2001; Boissier et al. 2001, 2003; Munoz-Mateos et al. 2007, 2009, 2011). This star formation history 
is parametrised with a polynom of the form:

\begin{equation}
\log SFR(t)_{secular} = a + b \log t + c t^{0.5}
\end{equation} 

\noindent
where $t$ is the time ellapsing after the first generation of stars are formed and the coefficients $a$, $b$, and $c$ for different rotational
velocities are given in Table \ref{buat}, the rotational velocity being the free parameter. 
A finer grid of $a$, $b$, and $c$ coefficients has been determined by interpolating the values given in Table \ref{buat} 
in the velocity range 40-360 km s$^{-1}$.

\begin{table}
\caption{Coefficients of the parametric star formation history. }
\label{buat}
{
\[
\begin{tabular}{cccc}
\hline
\noalign{\smallskip}
\hline
Velocity   &  a     & b	    & c      \\
km s$^{-1}$&        &       &        \\
\hline
40         & 4.73   & -0.11 & 0.79   \\
50         & 5.28   &  0.03 & 0.68   \\
60         & 5.77   &  0.16 & 0.57   \\
70         & 6.21   &  0.29 & 0.47   \\
80         & 6.62   &  0.41 & 0.36   \\
90         & 6.99   &  0.51 & 0.27   \\
100        & 7.34   &  0.61 & 0.18   \\
150        & 8.74   &  0.98 &-0.20   \\
220        & 10.01  &  1.25 &-0.55   \\
290        & 10.82  &  1.36 &-0.74   \\
360        & 11.35  &  1.37 &-0.85   \\
\noalign{\smallskip}
\hline
\end{tabular}
\]
}
\end{table}

Following this star formation history, in massive galaxies the bulk of the stars are formed at early epochs, 
while in dwarf systems the star formation activity is still rising. 
This parametrisation has been preferred to other analytical forms because its main parameter regulating the secular evolution of galaxies, 
the rotational velocity, is an observable available for the sample galaxies. This observable is crucial for constraining the 
secular evolution of the target galaxies independently from the spectrophotometric data, and thus its use reduces any possible degeneracy in the determination 
of the present star formation rate, quenching age and quenching factor of the target galaxies. 
To check whether this assumption does not introduce any systematic bias in the results, in Appendix A 
we perform the same analysis presented in this and in the following sections also levaing the rotational velocity as a free parameter, 
and we show that our results are reliable.  
The rotational velocity is derived from the HI line width $WHI$
measured using single dish observations and reported in Boselli et al. (2014a) through the relation:

\begin{equation}
{vel_{rot} = \frac{WHI}{2 \times \sin(incl)}}
\end{equation}

\noindent
where $incl$ is the inclination of the galaxy\footnote{The factor 2 takes into account the fact that $WHI$ is the total (receding plus approaching)
HI line width.}.
For galaxies with an inclination $incl$ $\leq$ 30$^o$, or galaxies without HI data, the rotational velocity is derived using the stellar mass Tully-Fisher 
relation\footnote{The stellar mass relation has been preferred to the baryonic Tully-Fisher relation just because molecular gas masses 
are not available for the low mass galaxies of the sample.} determined for the late-type galaxies of the HRS using the 
set of data given in Boselli et al. (2014a) (see Fig. \ref{TF}). 
This relation is (linear fit):

   \begin{figure}
   \centering
   \includegraphics[width=14cm]{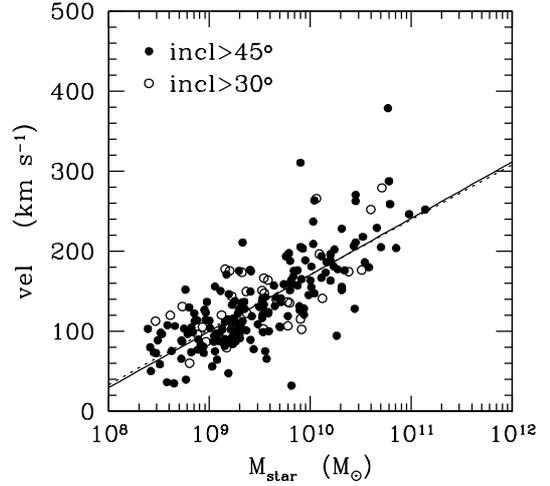}
   \caption{The rotational velocity vs. stellar mass relation derived for HRS galaxies with an inclination $incl$ $>$ 45$^o$ (filled dots) 
   and $incl$ $>$ 30$^o$ (filled and empty dots). The solid and dotted lines indicate the linear best fit to the data for galaxies with 
   $incl$ $>$ 45$^o$ and $incl$ $>$ 30$^o$, respectively.
 }
   \label{TF}%
   \end{figure}

\begin{eqnarray}
vel_{rot} ~[\rm{km~s^{-1}}] &=& 68.71\pm4.17 \times \log \it{M_{star}} ~[\rm{M_{\odot}}]\\
&& - 516.11\pm39.68 ~~(\rho = 0.75) \nonumber
\end{eqnarray}

\noindent
(where $\rho$ is the correlation coefficient), has been derived using 217 galaxies of the sample with available data\footnote{The relation does
not change significantly if the fit is restricted to a smaller sample of galaxies with $incl$ $>$ 45$^o$.}.
For the early-type galaxies of the sample, the rotational velocity is tentatively derived from 
the same stellar mass-rotational velocity relation derived for late-type galaxies (den Heijer et al. 2015). Although early-type systems do not follow
the same scaling relations than spirals, this assumption is justified by the fact that we want to test whether these typical 
cluster objects have been formed through the transformation of star forming systems after a gas stripping phenomenon.

To reproduce the quenching of the star formation activity of the Virgo cluster galaxies we use a formalism similar to the one proposed in Ciesla et al. (2016), i.e.
we apply an instantaneous truncation of the star formation history parametrised by the expression:

	\begin{equation}
	$$
	SFR(t) = \left\{ \begin{array}{ll}
	SFR(t)_{secular} &\mbox{ if $t_0$-$t$ $\geq$ QA} \\
	(1 - QF) \times SFR(t_0-QA)_{secular}  &\mbox{ if $t_0$-$t$ $<$ QA} 
	\end{array} \right.
	$$
	\end{equation}

   \begin{figure}
   \centering
   \includegraphics[width=9cm]{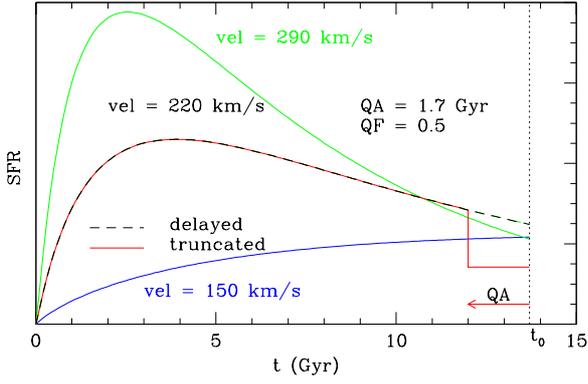}
   \caption{The parametric star formation history used in this work for a galaxy with a rotational velocity of 220 km s$^{-1}$ (black dashed line),
   290 km s$^{-1}$ (green solid line) and 150 km s$^{-1}$ (blue solid line).
   The black dashed line shows the delayed star formation history given in eq. (1), the red solid line the truncated one (eq. 4) for the case $vel$ = 220 km s$^{-1}$
   (adapted from Ciesla et al. 2016).
   }
   \label{SFH}%
   \end{figure}

\noindent
where $SFR(t)_{secular}$ is the star formation rate expected for a secular evolution (eq. 1), $QA$ is the quenching age,
$QF$ the quenching factor and $t_0$ the age of the universe at the present epoch, as depicted in Fig. \ref{SFH}.
This is obviously a crude representation of the physical process affecting galaxies. It is indeed known from observations and 
simulations that the gas stripping process is not instantaneous, but rather requires a certain time to be efficient.
Furthermore the duration of the stripping process strongly depends on the perturbing mechanism: it is very long (several Gyrs)
for starvation (Boselli et al. 2006) and harassment (Moore et al. 1998), where multiple flyby encounters of cluster galaxies are required, 
while it is relatively short for ram pressure. We prefer, however, this simple representation to a more physically motivated 
parametrisation such as the one proposed in Wetzel et al. (2013) because we want to reduce the number of free
parameters and thus limit any possible degeneracy in the derived quantities.

Hydrodynamical simulations consistently indicate that the total amount of gas of  
typical cluster galaxies is removed by ram pressure in $\lesssim$ 1.5 Gyr (Roediger \& Bruggen 2007; Tonnesen \& Bryan 2009). Observations and simulations, 
however, also indicate that most of the gaseous component is removed on much shorter timescales 
because the ram pressure stripping efficiency, which scales as $\rho$$V^2$ (Gunn \& Gott 1972, where $\rho$ is the density of the ICM and $V$ the
velocity of galaxies within the cluster), is at its maximum only when the galaxy is
crossing the core of the cluster. This is well depicted in Figure 3 of Vollmer et al. (2001) which indicates that the efficiency of 
ram pressure increases by a factor of $\simeq$ 10 on timescales $\lesssim$ 100 Myr. These timescales for gas stripping and quenching of the 
star formation activity of Virgo cluster galaxies are comparable to those derived with independent techniques on a dozen of well known Virgo cluster galaxies (see sect. 5).

The parametrisation given in eq. 4 to reproduce the truncation of the star formation activity of cluster galaxies 
is, however, extreme. To test how much the results obtained in the following analysis depend on this assumption
we also use as other extreme case a smooth constantly declining star formation history (see Appendix A). The main results of the
analysis do not change significantly.

\subsection{Spectral line indices}

\begin{figure}
   \centering
   \includegraphics[width=15cm]{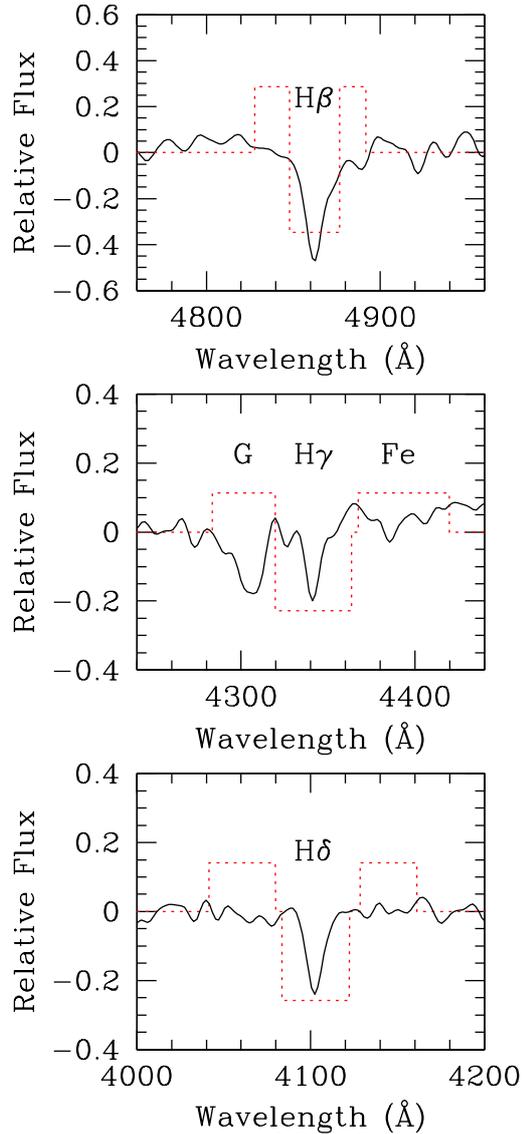}
   \caption{The transmissivity of the newly defined pseudo filters (red dotted line) centered on the age-sensitive H$\beta$, H$\gamma$, and H$\delta$
   lines is plotted as a function of wavelength and compared to the integrated spectrum of NGC 4569 once the emission lines are removed (black solid line; Boselli et al. 2013). 
 }
   \label{filters}%
   \end{figure}

To constrain the typical age of truncation of the star formation activity of the perturbed cluster galaxies 
we add several age-sensitive spectral indices. To be used and weighted as the other broad photometric bands 
in the CIGALE SED fitting code, these spectral indices must be defined as photometric bands.
The most widely used age-sensitive spectral indices present in the visible spectra of galaxies are the Balmer
absorption lines (H$\beta$, H$\gamma$, H$\delta$) and the $D_{4000}$ (Worthey 1994; Poggianti \& Barbaro 1997; 
Balogh et al. 1999; Kauffmann et al. 2003; Brinchmann et al. 2004). In the standard Lick/IDS system, the Balmer
indices are defined as equivalent widths. They cannot be directly used in CIGALE. We thus defined some pseudo filters
centered on the Balmer lines where the transmissivity is set positive on the stellar continuum on the wedges of the
absorption line and negative on the absorption line, as depicted in Fig. \ref{filters}. The integral of the transmissivity
on the blue and red stellar continua is set equal to that over the absorption line index, so that for flat spectra
the total flux within these pseudo filters is zero:

\begin{equation}
{\int_{blue_1}^{blue_2}{T({\lambda})d{\lambda}} + \int_{red_1}^{red_2}{T({\lambda})d{\lambda}} = - \int_{line_1}^{line_2}{T({\lambda})d{\lambda}}}
\end{equation}

\noindent
and

\begin{equation}
{\int_{line_1}^{line_2}{T(\lambda)d{\lambda}} = 1}
\end{equation}

\noindent
for the normalisation of the filter transmissivity,
where $blue_1$ and $blue_2$ and $red_1$ and $red_2$ stand for the limits at shorter and longer wavelengths where the continuum in the Lick/IDS indices is 
measured, while $line_1$ and $line_2$ the interval in wavelengths for the absorption line\footnote{The transmissivity of the filter is set negative on the absorption line just because the
CIGALE fitting code requires positive values in the different bands.}. The first condition is required to make the contribution of the stellar continuum
on the on and off bands the same. To be as consistent as
possible with the Lick/IDS definition, the continuum and the line emission are measured on the same intervals as in
Worthey \& Ottaviani (1997) and Worthey (1994) (wide filter definition, see Table \ref{Tabfilters}).\\

The $D_{4000}$ spectral index generally used in the literature (e.g. Balogh et al. 1999) is defined as a ratio of the flux 
measured in two spectral ranges at shorter and longer wavelengths than the spectral discontinuity. Being defined as a flux ratio, 
this index cannot be simulated with a similarly defined pseudo filter and thus it will not be used in this analysis.

\begin{figure}
   \centering
   \includegraphics[width=10cm]{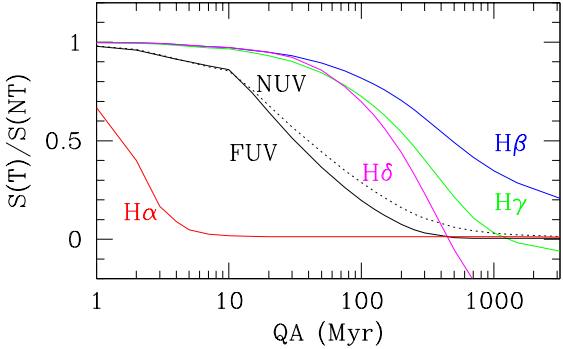}
   \caption{The expected variation of the flux density $S(T)$ for a truncated star formation history 
   in the $FUV$ and $NUV$ bands and in the newly defined
   LyC (H$\alpha$), H$\beta$, H$\gamma$, and H$\delta$ pseudo filters after a total quenching of the star formation
   activity (quenching factor $QF$ = 1) with respect to that of an unperturbed galaxy ($S(NT)$)
   as a function of the quenching age $QA$ for a model galaxy with properties given in
   Table \ref{Tabmock}. 
 }
   \label{modello}%
   \end{figure}

The H$\alpha$ emission, which is due to the emission of the gas of the ISM ionised by 
the UV photons produced in O and early-B massive stars ($\gtrsim$ 10 M$_{\odot}$; Kennicutt 1998b; Boselli et al. 2009), is sensitive to much
younger stellar populations ($\simeq$ 5 Myr; Fig. \ref{modello}). Population synthesis models provide the number of ionising photons $N_{\nu}$ for a 
given star formation history. We thus use the H$\alpha$ luminosities derived from narrow band imaging data to estimate the number 
of ionising photons using the relation (Osterbrock \& Ferland 2006):

\begin{equation}
{N_{\nu} = \frac{L(H\alpha)  ~[\rm{erg ~s^{-1}}]}{1.363 \times 10^{-12}}}
\end{equation}

\noindent
where $L(H\alpha)$ is the H$\alpha$ luminosity. Narrow band H$\alpha$+[NII] imaging data are first corrected for
[NII] contamination using the long slit spectroscopy of Boselli et al. (2013) (using the updated table given in Boselli et al. 2015).
We then make the hypothesis that all the ionising radiation ($\lambda$ $<$ 912 \AA) is absorbed by the gas, or in other words that 
the escape fraction is zero and that the ionising photons do not contribute to the heating of dust ($f$=1).
Narrow band H$\alpha$+[NII] imaging data are also corrected for dust attenuation. This is done using the Balmer decrement whenever possible, 
otherwise using standard recipes based on monochromatic 22 $\mu$m WISE data as described in Boselli et al. (2015).
CIGALE requires flux densities measured within a filter bandpass. For this reason we defined another pseudo filter 
($Ly_C$) to characterise the ionising radiation with a positive and constant transmissivity for $\lambda$ $<$ 912 \AA:

\begin{equation}
{\int_{0}^{912 \AA}{T(\lambda)d{\lambda}} = 1}
\end{equation}

\noindent
Since CIGALE requires a flux density in this photometric band, the number of ionising photons must be transformed into mJy.
This is done using a calibration that we derived on the mock catalogue (see sect. 4.4)  
by measuring on the spectral energy distribution of the simulated galaxies the flux density within this pseudo filter  
and comparing it to the number of ionising photons given by the population synthesis models:

\begin{equation}
{LyC ~\rm{[mJy]} = \frac{1.07 \times 10^{-37} \times \it{L(H\alpha)} ~\rm{[erg ~s^{-1}]}}{ \it{D}^2 ~\rm{[Mpc]}}} 
\end{equation}

\noindent
or equivalently:

\begin{equation}
{LyC ~\rm{[mJy]} =  1.079 \times 10^4 \frac{ \it{SFR_{Salp}} ~\rm{[M_{\odot} yr^{-1}]}}{\it{D}^2 ~\rm{[Mpc]}}}
\end{equation}


Figure \ref{modello} shows how the flux within these newly defined pseudo filters and in the GALEX $FUV$ and $NUV$ age-sensitive photometric bands 
changes when the star formation activity of a typical star forming model
galaxy with parameters given in Table \ref{Tabmock} is abruptly truncated. The Balmer absorption line indices are sensitive to stellar populations of intermediate ages
and are thus indicated to characterise the quenching age for truncations $\simeq$ 500 Myr old. The $FUV$ and $NUV$ flux densities are sensitive to 
relatively younger ages ($\simeq$ 100-200 Myr), while the number of Lyman continuum photons derived from H$\alpha$ imaging data to much 
younger stellar populations ($\lesssim$ 5 Myr; Boquien et al. 2014). The determination of the uncertainty on the flux in the pseudo filters is made difficult
by several factors (uncertainty on the normalisation, noise in the stellar continuum, electronic noise in the spectrum). 
Similarly, the uncertainty on the number of ionised photons derived from narrow band H$\alpha$ imaging data is hardly quantifiable because 
of the different steps required to transform observables (H$\alpha$+[NII] fluxes) with their own uncertainties to physical quantities 
([NII] contamination, dust attenuation correction...). There are also large uncertainties in the stellar population synthesis 
models on the intensity of the Lyman continuum photons (Levesque et al. 2012). We thus assume a conservative uncertainty of 15\% in all the pseudo filters,
but recall that the CIGALE code adds a 10\% ~ uncertainty as in all other bands. The same statistical weight is applied to all photometric bands in the fitting procedure (including the newly defined pseudofilters)

\begin{table}
\caption{Pseudo filters definitions. }
\label{Tabfilters}
{
\[
\begin{tabular}{cccc}
\hline
\noalign{\smallskip}
\hline
Filter		&  $blue_1-blue_2$	& $line_1-line_2$	& $red_1-red_2$       \\
\hline
H$\beta$	& 4827.875-4847.875	& 4847.875-4876.625	& 4876.625-4891.625	\\
H$\gamma$	& 4283.50-4319.75	& 4319.75-4363.50	& 4367.25-4419.75	\\
H$\delta$	& 4041.60-4079.75	& 4083.50-4122.25	& 4128.50-4161.00	\\
\noalign{\smallskip}
\hline
\end{tabular}
\]
}
\end{table}


   \begin{figure}
   \centering
   \includegraphics[width=10cm]{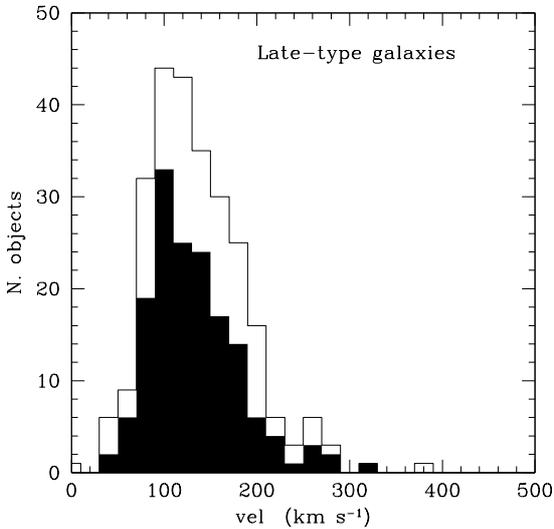}
   \caption{Distribution of the rotational velocity of all the late-type galaxies of the sample (empty histogram) and of
   those objects with data in the newly defined pseudo filter bands. 
    }
   \label{distvel}%
   \end{figure}

Figure \ref{distvel} gives the rotational velocity distribution for the late-type galaxies of the sample, and indicates that
the critical parameters necessary for constraining the recent star formation activity (the pseudo filter bands as well as the $FUV$ and
$NUV$ bands) are available for most of them. It also shows that the range in rotational velocity of the HRS galaxies is within 
the one covered by the star formation history given in eq. 1 and in Table \ref{buat}.

\subsection{Mock catalogue of simulated galaxies}

To test the solidity of this approach, we first generate a grid of simulated galaxies and extract their SED
in the same photometric and spectroscopic bands analysed in this work by varying different input parameters
in the CIGALE code, as listed in Table \ref{Tabmock}. These different parameters have been chosen to sample in a relatively
uniform way the expected parameter space for the HRS galaxies in terms of star formation history, dust attenuation and dust
properties. Since the most critical parameters for the following analysis are the present day star formation rate, the star formation
history, the quenching factor and the quenching age of the perturbed galaxies, these parameters have been sampled in narrower bins
than the other variables in the simulated galaxies. The resulting catalogue of simulated SED includes 
282528 objects.\\
We then produced a mock catalogue by artificially introducing noise in the simulated SED according to the 
typical error in the different photometric bands. We then fitted this simulated mock catalogue of SED
and made a PDF analysis of the most critical variables. 
To avoid edge effects in the PDF analysis, we analysed a wider range in the parameter space than in the one used to construct
the SEDs. This has been done only for those variables analysed in
this work, i.e. the quenching factor and the quenching age, as described in Table \ref{Tabmock}. Given the physical limit of the two variables 
(the quenching age must be $\geq$ 0 and the quenching factor cannot be $>$ 1), the PDF cannot be sampled symmetrically. We also
sampled negative quenching factors which might be representative of a recent burst of star formation (see Table \ref{Tabmock}).\\

\begin{table*}
\caption{Input parameters used to create the mock catalogue as defined in sec. 4.1. }
\label{Tabmock}
{
\[
\begin{tabular}{ccc}
\hline
\noalign{\smallskip}
\hline
Parameter	&  value			& Units         \\
\hline
Pop.Synth.Mod.	& Bruzual \& Charlot (2003)	& 	 	\\
Dust model	& Draine \& Li (2007)		&		\\
IMF	 	& Salpeter			&		\\
Metallicity	& \textbf{0.02}				&		\\
Velocity 	& 40, 80, 120, 160, \textbf{200}, 240, 280, 320, 360& km s$^{-1}$\\
$QF$		& -0.4, -0.2, 0, 0.2, 0.4, 0.6, 0.7, 0.8, 0.9, 0.95, \textbf{1}&		\\
$QA$		& 0, 5, 10, 20, 50, 100, 150, 200, 250, 300, 500, 700, 1000, 1500 & Myr	\\
$E(B-V)_{young}$& 0.05, 0.1, 0.2,\textbf{0.4}	&		\\
$E(B-V)_{old}$	& \textbf{0.44}			&		\\
$Q_{PAH}$	& 0.47, 2.50, \textbf{4.58}, 6.63	&		\\
$U_{min}$	& 0.1, 0.5, \textbf{2.0}, 4.0, 8.0, 20	&		\\
$\alpha$	& \textbf{2.5}			&		\\
$\gamma$	& 0.01, \textbf{0.03}, 0.1	&		\\
\noalign{\smallskip}
\hline
\end{tabular}
\]
Note: The parameters for the model galaxy used to trace the variation of the flux densities in the pseudo filters after a total quenching
of the star formation activity shown in Fig. \ref{modello} are given in boldface.
}
\end{table*}

   \begin{figure}
   \centering
   \includegraphics[width=9cm]{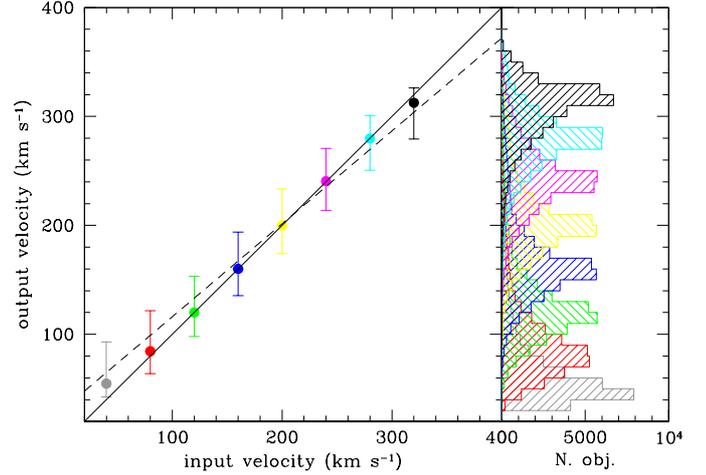}
   \caption{Relationship between the input rotational velocity of the simulated galaxies and the output velocity derived from the PDF
   analysis of the mock SED (left panel). Dots are median values while error bars are the 16\% and 84\% quartiles of the distribution.
   The solid line shows the 1:1 relationship, while the dashed line the linear best fit to the data. The right panel shows the
   distribution of output velocities derived from the PDF analysis for a given value of the input velocity for the simulated
   galaxies. }
   \label{vel}%
   \end{figure}

   \begin{figure}
   \centering
   \includegraphics[width=9cm]{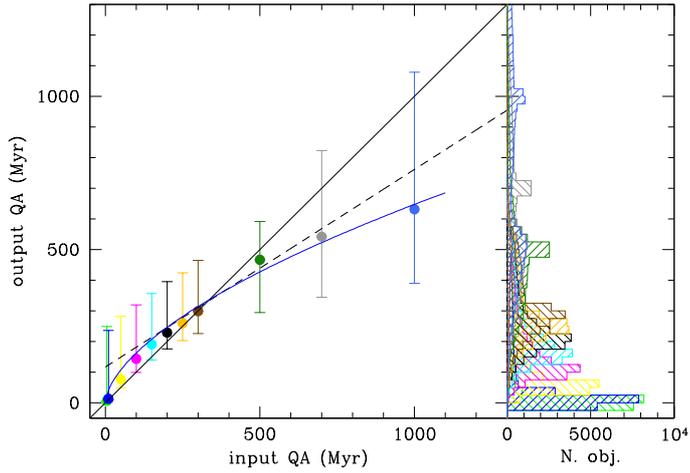}
   \caption{Relationship between the input quenching age of the simulated galaxies and the output quenching age derived from the PDF
   analysis of the mock SED (left panel). Dots are median values while error bars are the 16\% and 84\% quartiles of the distribution.
   The solid line shows the 1:1 relationship, while the dashed line the linear best fit to the data, the blue solid line a second degree polynomial fit. 
   The right panel shows the
   distribution of output quenching ages derived from the PDF analysis for a given value of the input quenching age for the simulated
   galaxies. }
   \label{QA}%
   \end{figure}

   \begin{figure}
   \centering
   \includegraphics[width=9cm]{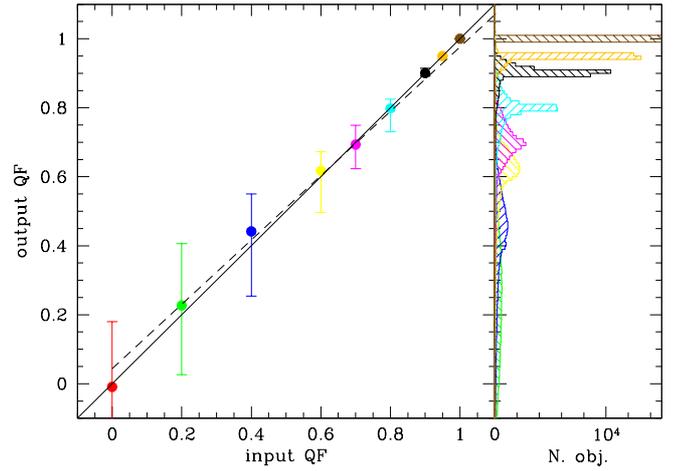}
   \caption{Relationship between the input quenching factor of the simulated galaxies and the output quenching factor derived from the PDF
   analysis of the mock SED (left panel). Dots are median values while error bars are the 16\% and 84\% quartiles of the distribution.
   The solid line shows the 1:1 relationship, while the dashed line the linear best fit to the data. The right panel shows the
   distribution of output quenching factor derived from the PDF analysis for a given value of the input quenching factor for the simulated
   galaxies. }
   \label{QF}%
   \end{figure}

\begin{figure}
   \centering
   \includegraphics[width=9cm]{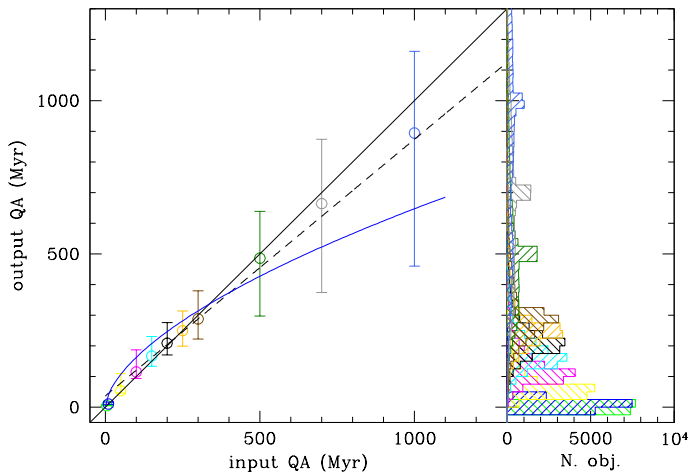}
   \caption{Relationship between the input quenching age of the simulated galaxies and the output quenching age derived from the PDF
   analysis of the mock SED restricted to galaxies with an input quenching factor QF$>$ 0.5 (left panel). 
   Dots are median values while error bars are the 16\% and 84\% quartiles of the distribution.
   The solid line shows the 1:1 relationship, while the dashed line the linear best fit to the data, the blue solid line a polynomial fit
   derived for the whole mock sample shown in Fig. \ref{QF}. 
   The right panel shows the
   distribution of output quenching ages derived from the PDF analysis for a given value of the input quenching age for the simulated
   galaxies. }
   \label{QAgeQF05}%
   \end{figure}

Figures \ref{vel}, \ref{QA}, and \ref{QF} show the relationship between the input parameters used to define the star
formation history of the simulated galaxies and the distribution of output parameters derived from SED fitting
of the mock catalogue. The analysis of these Figures reveals that:\\
1) The rotational velocity, which is the parameter regulating the secular evolution of galaxies and thus their long term star 
formation history, is well measured in the PDF analysis of the mock catalogue. \\
2) The quenching age extracted from the PDF analysis matches quite well the input values for quenching ages $\lesssim$ 500 Myr, while
it systematically underestimates the input value for older quenching ages (by $\sim$ a factor of 2 for a quenching age of 1 Gyr). 
The uncertainty on the derived parameter increases with increasing input quenching ages. \\
3) The quenching factor extracted from the PDF analysis is tightly connected to the input value. The uncertainty on its measure,
however, increases with decreasing quenching factor. \\
Point 2) and 3) can be understood if we consider that whenever the truncation occurred at early epochs (large $QA$) and/or the quenching factor is 
small, the imprints of these effects on the SED are hardly distinguishable from those due to possible variations in the secular evolution of galaxies.
Indeed, if we limit for instance the comparison of the input quenching age of the simulated galaxies and the output quenching age derived from the PDF
analysis of the mock SED to those galaxies with $QF$ $>$ 0.5, thus to those objects where the truncation of the star formation activity 
is significant, we obtain significantly better results (see Fig. \ref{QAgeQF05}).

We already analysed the reliability of using a SED fitting analysis to study short term variations in the star formation 
activity of perturbed galaxies in Ciesla et al. (2016). The results of our analysis are directly comparable since based on the same 
sample of galaxies and on the same set of photometric data. 
Ciesla et al. (2016) have shown that, whenever UV data are available, the SED fitting analysis gives a reliable 
estimate of the quenching factor, but fails to estimate the quenching age and $\tau_{main}$, the typical timescale characterising 
the secular evolution of galaxies. As discussed in Ciesla et al. (2016), this is probably due to a degeneracy between the free 
parameters characterising the parametrised star formation history. To overcome this problem and capitalise on the promising results 
obtained in Ciesla et al. (2016) we adopt here a different star formation history where the secular evolution of galaxies 
is not a free parameter as in Ciesla et al. (2016; the $\tau_{main}$ parameter) but rather it is observationally constrained from the
rotational velocity of the galaxy as derived from HI data (see sec. 4.2). The number of free parameters necessary to parametrise the star
formation history is thus reduced from three to two. At the same 
time we use a significantly larger number of age sensitive photometric bands
(LyC, $FUV$, $NUV$, H$\beta$, H$\gamma$, H$\delta$) than in Ciesla et al. (2016) ($FUV$, $NUV$). In particular, the
introduction of the number of ionising photons derived from the H$\alpha$ emission, which are produced in young ($\lesssim$ 10$^7$
yr) and massive ($M$ $\gtrsim$ 10 M$_{\odot}$) stars, is critical for constraining recent episodes of star formation. These improvements 
reduce the degeneracy between the fitted parameters and  have a significant impact on the reliability of their determination.
The comparison of Figures \ref{QA} and \ref{QF} with Figures 4 and 5 of Ciesla et al. (2016), indeed, indicates a remarkable 
increase in the accuracy in the quenching parameters derived in the present work. \\



\section{SED fitting of representative galaxies}

A further test in the reliability of the output parameters of our SED fitting code can be done by comparing
them to those derived with independent techniques on a few well known galaxies of the sample. A dozen of Virgo cluster 
galaxies of the HRS, indeed, have been studied in great detail using a combination of spectro-photometric and kinematic data
with models tuned to reproduce the interaction of galaxies with the hostile cluster environment.
 
\subsection{NGC 4569}

We first test the reliability of our fitting procedure using NGC 4569, the brightest late-type galaxy of the Virgo cluster.
This galaxy is undergoing a ram pressure stripping event as revealed by a dynamically perturbed velocity field of the gaseous component (Vollmer et al.
2004), a truncated gaseous and stellar disc in the young stellar populations (H$\alpha$, $FUV$, $NUV$), while 
a normal extended disc in the old stellar populations (Boselli et al. 2006), and a $\sim$ 100 kpc extended tail of ionised gas formed during the
interaction with the hot and dense intracluster medium (Boselli et al. 2016). The dynamical models of Vollmer et al. (2004) indicate that the
gas stripping process took place $\sim$ 300 Myr ago. Using tuned models of galaxy evolution especially tailored to take into account the effects of
ram pressure stripping, Boselli et al. (2006) were able to reproduce the observed radial truncation in the gaseous disc and in the
stellar disc at different wavelengths if the peak of the stripping process occurred $\sim$ 100 Myr ago. By studying the stellar population properties 
of the outer disc of NGC 4569 using IFU spectroscopic data, Crowl \& Kenney (2008) concluded that the star formation activity of NGC 4569 ended outside the
truncation disc $\sim$ 300 Myr ago. We thus have consistent and independent indications that this galaxy was perturbed recently (100-300 Myr). The
models of Boselli et al. (2006) also indicate that the activity of star formation of this galaxy was reduced by 95\% ($QF$ = 0.95).

\begin{figure}
   \centering
   \includegraphics[width=9cm]{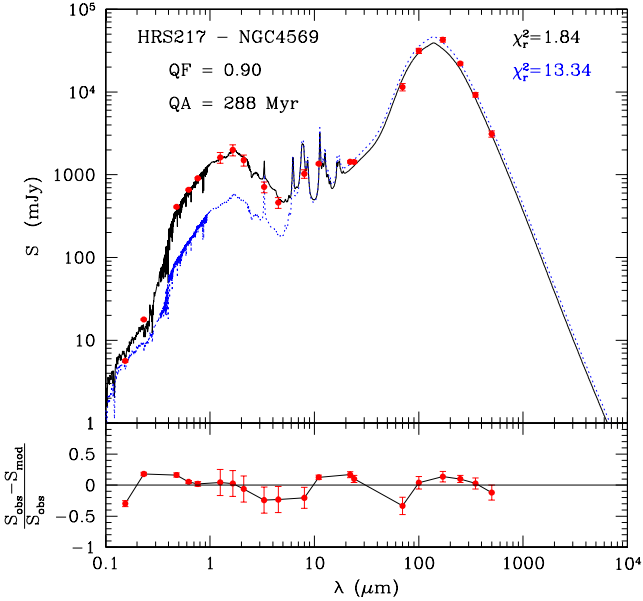}
   \caption{The observed UV to FIR SED of NGC 4569 (HRS 217) (red filled dots) is compared to the best fitted model derived by CIGALE using a 
   truncated star formation history (black solid line) or a normal delayed star formation history (blue dotted line) (upper panel). 
   The normalised difference between the observed and model value in the different photometric bands 
   is plotted vs. lambda in the lower panel (for a truncated star formation history).
    }
   \label{SED}%
   \end{figure}

\begin{figure}
   \centering
   \includegraphics[width=9cm]{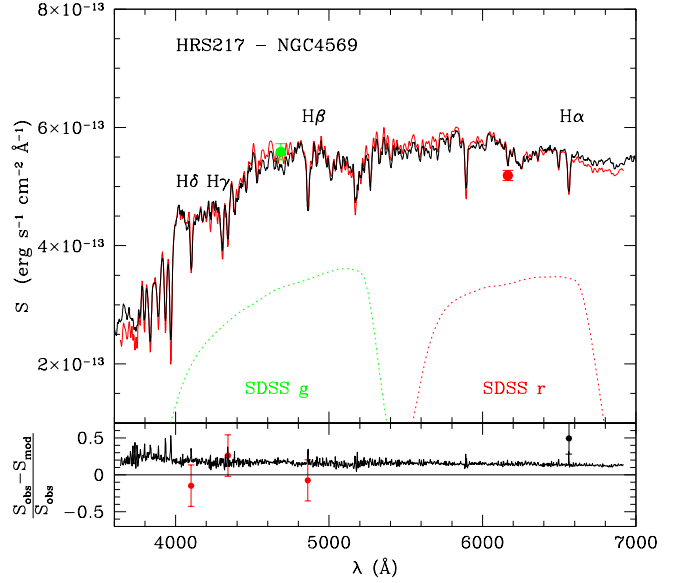}
   \caption{The integrated spectrum of NGC 4569, with a resolution of $R$ $\sim$ 1000, corrected for line emission (red line), normalised to the flux density in the $g$
   band (green filled dot), is compared to the best fitted model derived by CIGALE using a 
   truncated star formation history (black solid line; the model spectrum is shifted on the Y-axis by a factor of $+$8$\times$10$^{-14}$ to 
   match the observed spectrum). The red filled dot shows the $r$-band photometric point, while the transmissivity of the SDSS $g$ and $r$ bands 
   is traced by the green and red dotted curves (upper panel). The normalised difference between the two spectra 
   is plotted vs. lambda in the lower panel. The red filled dots indicate the normalised difference between the values of the H$\beta$, H$\gamma$, and H$\delta$ 
   pseudo filters and those measured on the best fitted model. The black filled dot at $\lambda$ 6563 \AA ~shows the normalised difference between the observed Lyman continuum
   pseudo filter, derived from the extinction corrected H$\alpha$ image of the galaxy, and the Lyman continuum flux measured on the best fitted model.
    }
   \label{spettro}%
   \end{figure}

The observed UV to FIR SED of NGC 4569 is well reproduced by CIGALE using a truncated star formation history, while the fit does not match the photometric 
data in the UV to NIR bands if a simple delayed star formation history is adopted (Fig. \ref{SED}). The match between
the best model and the observations is evident also at high resolution in the optical domain, where the truncated model can be directly compared to the 
normalised observed spectrum (Fig. \ref{spettro}). Despite a systematic shift in the model vs. observations already present in the UV to FIR SED in the $g$ band 
used for the normalisation of the integrated spectrum, there is a very good agreement in the high resolution absorption features, and in
particular in the age sensitive Balmer absorption lines used in the fit.
 
Our SED fitting code gives $QA$ = 288$\pm$23 Myr and a $QF$ = 0.90$\pm$0.05, values very consistent with those derived with independent techniques. 
NGC 4569 also belongs to the SINGS sample of galaxies (Kennicutt et al. 2003) and thus have an integrated spectrum taken by Moustakas et al. (2010). If we 
combine the photometric data to the spectroscopic dataset of Moustakas et al. (2010) we derive very similar values for the quenching age ($QA$ = 299$\pm$15 Myr)
and the quenching factor ($QF$=0.90$\pm$0.05), indicating that the derived values are robust vs. the use of different spectroscopic datasets.

\subsection{Other galaxies}

\begin{figure}
   \centering
   \includegraphics[width=10cm]{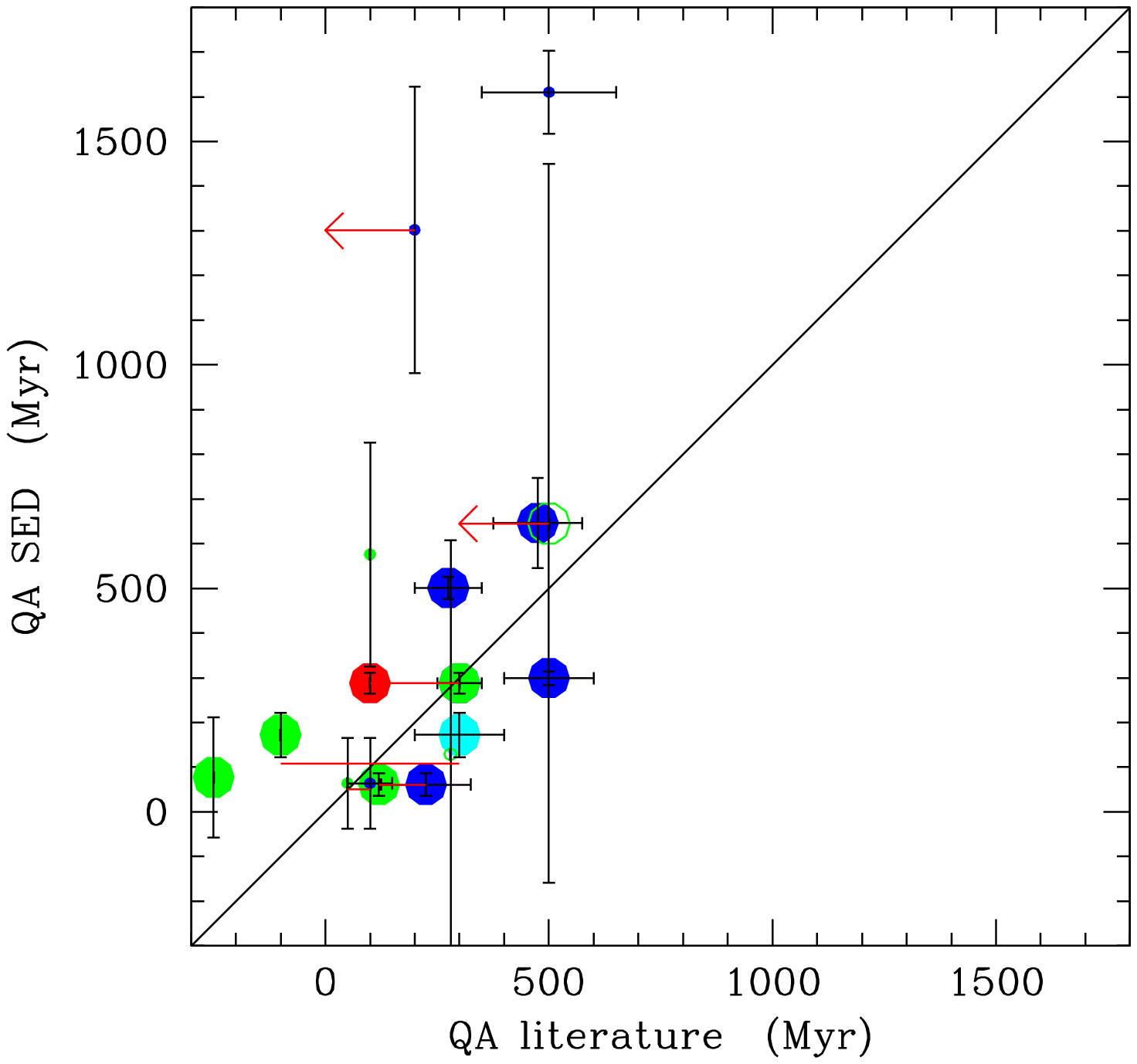}
   \caption{Relationship between the quenching age derived using our SED fitting technique and that derived using multizone 
   chemo-spectrophotometric models of galaxy evolution for NGC 4569 (red dot), dynamical models based on the HI and CO gas kinematics (green dots),
   IFU spectroscopy of the outer disc of some cluster galaxies (blue dots), and photometric data compared to population synthesis models (cyan dots). 
   Filled symbols are used for galaxies with a quenching
   factor $QF$ $>$ 0.5, empty dots for those with $QF$ $\leq$ 0.5. Large symbols are for galaxies with a reduced $\chi_r^2$ $\leq$ 3, 
   small symbols for  $\chi_r^2$ $>$ 3. The red lines connecting two points are used to indicate those galaxies with two independent 
   quenching ages available in the literature. The red arrows indicate upper limits to the quenching age given in the literature. The data used for this
   plot are given in Table \ref{Tabknown}. The solid diagonal line indicates the 1:1 relationship. 
    }
   \label{known}%
   \end{figure}

The same comparison can be extended to a few other galaxies with estimates of the quenching age from dynamical models
or from IFU spectroscopic data of the outer discs (see Fig. \ref{known} and Table \ref{Tabknown}). We recall that the timescales derived using dynamical models 
indicate the epoch when the gaseous component started to be perturbed. This might not coincide with the beginning of the decrease of
the star formation activity, which generally occurs after the gas has been stripped (Boselli et al. 2008a; Crowl \& Kenney 2008). Furthermore,
the analytical definition of the star formation history used in this work does not allow negative quenching ages as dynamical models do (the
gas stripping process starts before the peak of the ram pressure stripping process has been reached). 
Despite the large uncertainty in the plotted parameters, overall there is a reasonable agreement between the quenching ages derived using different 
techniques when the comparison is limited to those objects with a good quality fit (reduced $\chi_r^2$ $\leq$ 3).

\begin{table*}
\caption{Quenching parameters derived using other techniques for some HRS galaxies in the Virgo cluster. }
\label{Tabknown}
{
\[
\begin{tabular}{ccccccccc}
\hline
\noalign{\smallskip}
\hline
HRS &	NGC/IC  & vel	    &  QA           &    QF         & $\chi_r^2$	& age                 &  ref  &  method\\   
    &           &km s$^{-1}$& Myr           &               & 			& Myr                 &       &        \\
(1) &  (2)      & (3)       & (4)           &    (5)        & (6)               & (7)    &  (8)  & (9) \\    
\hline
102	& 4254  & 175	    &  129 $\pm$ 478  &   -0.06$\pm$ 0.36& 3.14		& 280		 & 8	 & dyn  \\    
124	& 4330  & 120	    &  172 $\pm$ 50   &   0.89 $\pm$ 0.04& 1.09		&-100		 & 3	 & dyn  \\ 
124	& 4330  & 120	    &  172 $\pm$ 50   &   0.89 $\pm$ 0.04& 1.09		& 200-400	 & 11	 & phot  \\    
144	& 4388  & 180	    &  61  $\pm$ 25   &   0.74 $\pm$ 0.06& 1.00		& 225$\pm$100   & 6 	 & spec \\    
144	& 4388  & 180	    &  61  $\pm$ 25   &   0.74 $\pm$ 0.06& 1.00		& 120   	 & 9	 & dyn  \\    
149	& 4402  & 120	    &  1302$\pm$ 320  &   0.90 $\pm$ 0.04& 3.23		& 200$^a$ 	 & 6	 & spec \\     
156	& 4419  & 95	    &  1610$\pm$ 75   &   0.95 $\pm$ 0.05& 6.15		& 500$\pm$150   & 6 	 & spec \\    
159	& 4424  & 32	    &  501 $\pm$ 25   &   0.95 $\pm$ 0.05& 2.81		& 275$\pm$75	 & 6	 & spec \\    
163	& 4438  & 130	    &  576 $\pm$ 250  &   0.95 $\pm$ 0.05& 3.66		& 100		 & 4	 & dyn  \\    
172	&IC3392 & 100	    &  299 $\pm$ 15   &   1.0  $\pm$ 0.05& 2.90		& 500$\pm$100	 & 6 	 & spec \\    
190	& 4501  & 290	    &  77  $\pm$ 135  &   0.81 $\pm$ 0.09& 2.88		& -250		 & 7	 & dyn  \\    
197	& 4522  & 105	    &  64  $\pm$ 101  &   0.80 $\pm$ 0.04& 3.01		& 50		 & 5	 & dyn  \\    
197	& 4522  & 105	    &  64  $\pm$ 101  &   0.80 $\pm$ 0.04& 3.01		& 100$\pm$50	 & 6	 & spec \\    
217	& 4569  & 220	    &  288 $\pm$ 23   &   0.90 $\pm$ 0.05& 1.84		& 100		 & 1	 & model  \\    
217	& 4569  & 220	    &  288 $\pm$ 23   &   0.90 $\pm$ 0.05& 1.84		& 300		 & 2	 & dyn  \\    
217	& 4569  & 220	    &  288 $\pm$ 23   &   0.90 $\pm$ 0.05& 1.84		& 300$\pm$50	 & 6	 & spec \\    
221	& 4580  & 105	    &  646 $\pm$ 101  &   0.95 $\pm$ 0.05& 1.52		& 475$\pm$100   & 6 	 & spec \\    
247	& 4654  & 175	    &  645 $\pm$ 804  &   0.40 $\pm$ 0.21& 2.39		& 500$^a$ 	 & 10	 & dyn  \\     
\noalign{\smallskip}
\hline
\end{tabular}
\]
Column 1: HRS name; 
Column 2: NGC/IC name; 
Column 3; rotational velocity used in the CIGALE fit; 
Column 4: $QA$ and error derived using CIGALE; 
Column 5: $QF$ and error derived using CIGALE; 
Column 6: reduced $\chi_r^2$;
Column 7: age of the interaction derived in the literature; 
Column 8: References: 
1: Boselli et al. (2006);
2: Vollmer et al. (2004);
3: Vollmer et al. (2012);
4: Vollmer et al. (2009);
5: Vollmer et al. (2006);
6: Crowl \& Kenney (2008);
7: Vollmer et al. (2008a);
8: Vollmer et al. (2005);
9: Vollmer \& Huchtmeier (2003);
10: Vollmer (2003); 
11: Abramson et al. (2011);
Column 8: method used to derive the age of the interaction: $dyn$ stands for dynamical models based on the HI and/or CO data; $spec$ for 
IFU spectroscopy of the outer stellar disc; $model$ for multizone-chemospectrophotometric models of galaxy evolution; $phot$ derived using 
UV-to-optical photometry.\\
Notes: $a$ = upper limit.
}
\end{table*}

\section{SED fitting of the HRS}



\begin{table*}
\caption{Input parameters used to fit the HRS galaxies. }
\label{Tabfit}
{
\[
\begin{tabular}{ccc}
\hline
\noalign{\smallskip}
\hline
Parameter	&  value			& Units         \\
\hline
Pop.Synth.Mod.	& Bruzual \& Charlot (2003)	& 	 	\\
Dust model	& Draine \& Li (2007)		&		\\
IMF	 	& Salpeter			&		\\
Metallicity	& 0.02				&		\\
Velocity 	& 40-360 in step of 10 		& km s$^{-1}$   \\
$QF$		& -0.4, -0.2, 0.2, 0.4, 0.6, 0.7, 0.8, 0.9, 0.95, 1&		\\
$QA$		& 0, 5, 10, 20, 50, 100, 150, 200, 250, 300, 400, 500, 600, 700, 800, 900, 1000, 1250, 1500, 2000, 3000 & Myr	\\
$E(B-V)_{young}$& 0.05, 0.1, 0.2, 0.4		&		\\
$E(B-V)_{old}$	& 0.44				&		\\
$Q_{PAH}$	& 0.47, 2.50, 4.58, 6.63, 7.32	&		\\
$U_{min}$	& 0.1, 0.5, 2.0, 4.0, 8.0, 20	&		\\
$\alpha$	& 2.0, 2.5, 3.0			&		\\
$\gamma$	& 0.01, 0.03, 0.1		&		\\
\noalign{\smallskip}
\hline
\end{tabular}
\]
Note: The input parameters used in the analysis of the HRS span a wider range than those used to create the mock galaxies to avoid border effects.}
\end{table*}

Since we want to focus our
analysis on the quenching phenomenon, the model SED used in the following analysis are constructed using for each galaxy 
the appropriate rotational velocity as derived from kinematical data (or from the Tully-Fisher relation as explained in sect. 4.2
whenever the former is not available). The grid of parameters used to fit the observed data are given in Table \ref{Tabfit}.

   \begin{figure*}
   \centering
   \includegraphics[width=15cm]{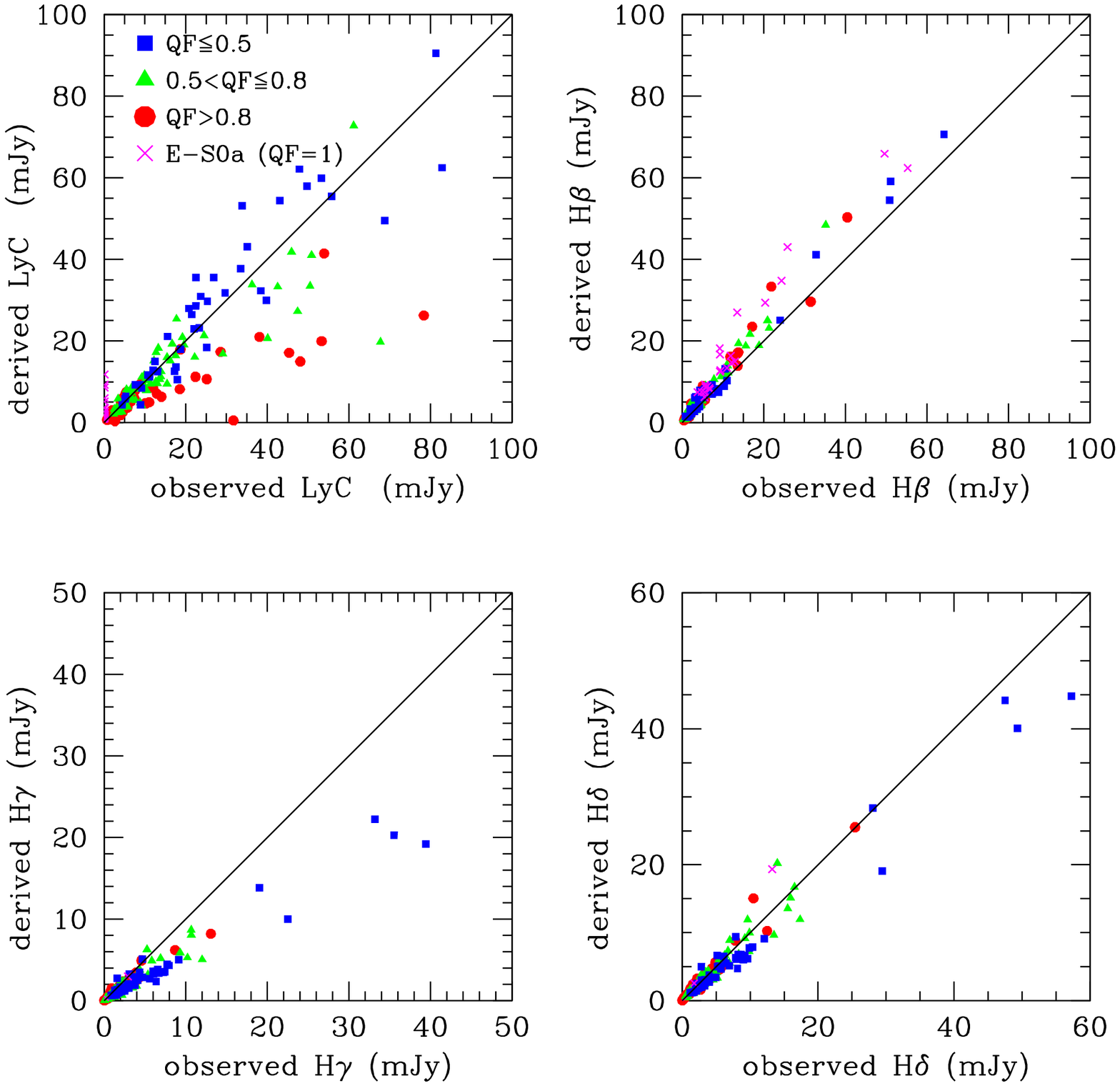}
   \caption{Relationship between the flux densities measured within the four different pseudo filters defined in this work on the best fitted model  
   and those measured on the observed spectra. The solid line indicates the 1:1 relation. 
   Blue filled squares, green triangles, red circles and magenta crosses are for galaxies where 
   the derived quenching factor is $QF$ $\leq$ 0.5, 0.5$<$ $QF$ $\leq$ 0.8, $QF$ $>$ 0.8, and early-type galaxies (all with $QF$=1), respectively. 
    }
   \label{pseudo}%
   \end{figure*}

Figure \ref{pseudo} shows the relationship between the flux densities in the four pseudo filters derived from the PDF analysis 
vs. the one measured on the observed spectra. The correlation between the two variable is excellent in the H$\beta$ and H$\delta$
bands, while the scatter increases in H$\gamma$ and particularly in the Lyman continuum derived from the H$\alpha$ data.
The low scatter in the H$\beta$ and H$\delta$ relations is probably due to the fact that both variables are sensitive to stellar 
populations of young-to-intermediate age (see Fig. \ref{modello}), already well constrained by the FUV-to-far-IR photometric bands.
The systematic difference observed in the H$\gamma$ pseudo filter might result from a possible contamination of the Fe and G bands 
in the stellar continuum (see Fig. \ref{filters}). The large scatter in the Lyman continuum pseudo filter is probably due to the 
fact that this is the only photometric band sensitive to very young ($\lesssim$ 10$^7$ yr) stellar populations. It is thus the most critical
(and unique) band for constraining recent variations in the star formation activity of the galaxies. Indeed, the 
correlation between the two variables is excellent whenever $QF$ $\leq$ 0.5 (mainly unperturbed objects), while there is a systematic
difference for the most perturbed  galaxies ($QF$ $>$ 0.8). The systematic variation of the correlation with $QF$ seen in 
Fig. \ref{pseudo} might be an artefact due to the very simplistic parametrisation of the star formation law used to 
describe the quenching phenomenon (eq. 4). Another possible origin of the observed scatter in the Lyman continuum pseudo filter can
be the fact that this observable is derived from H$\alpha$+[NII] imaging data once corrected for [NII] contamination 
using integrated spectroscopy and for dust attenuation using the Balmer decrement whenever possible, or the 22 $\mu$m emission from WISE
otherwise (Boselli et al. 2015). All these corrections are quite uncertain.

\section{Analysis}

To limit the uncertainty in the determination of the quenching age and quenching factor parameters, 
the analysis presented in this and in the following sections is restricted to the late-type galaxies of the HRS with available spectroscopic,
$FUV$, $NUV$, and H$\alpha$ imaging data, the last opportunely corrected for dust attenuation as described in the previous sections.
We recall that for all these galaxies photometric data are also available in the $gri$ from SDSS, $JHK$ from 2MASS, 
11 and 22 $\mu$m from WISE, 100 and 160 $\mu$m from PACS and 250, 350, 500 $\mu$m from SPIRE, while $\textit{Spitzer}$ data at 8, 24, and 70 $\mu$m for a large fraction of the
sample (see Table \ref{Tabunc}). 
We also analyse 22/61 of the early-type galaxies of the HRS with the same set of spectro-photometric data, with the exception of 
H$\alpha$ imaging data, but for which an upper limit to the H$\alpha$ emission has been derived from the spectra.
As stated in sect. 2, this sample includes galaxies in different density regions, from the core of the Virgo cluster
to small groups, binary systems and relatively isolated objects in the field (Boselli et al. 2010a).

\begin{figure}
   \centering
   \includegraphics[width=10cm]{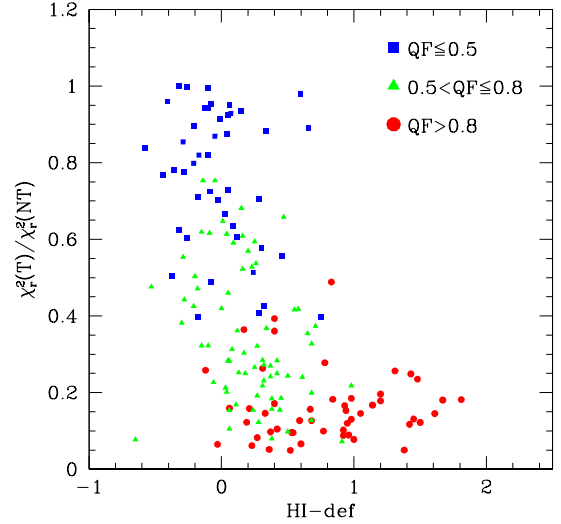}
   \caption{Relationship between the reduced $\chi^2_r$ ratio derived using a truncated ($\chi^2_r(T)$)
   to a non truncated ($\chi^2_r(NT)$) star formation history and the HI-deficiency parameter. Blue filled squares, green triangles, 
   and red circles are for late-type galaxies with a quenching factor $QF$ $\leq$ 0.5, 
   0.5$<$ $QF$ $\leq$ 0.8, and $QF$ $>$ 0.8, respectively. 
    }
   \label{chiv}%
   \end{figure}

We first test whether the use of a truncated star formation history increases the quality of the SED fitting procedure
of the perturbed Virgo cluster galaxies as in the case of NGC 4569 (see Fig. \ref{SED}). To do that, we plot the relationship 
between $\chi^2_r(T)$/$\chi^2_r(NT)$, where $\chi^2_r(T)$ is the reduced chi-square derived using a truncated star formation 
history (eq. 4) and  $\chi^2_r(NT)$ the one derived using an unperturbed star formation history (eq. 1), 
and the HI-deficiency parameter (Fig. \ref{chiv}), here taken as proxy for the perturbation.
We recall that the typical HI-deficiency of isolated galaxies is $HI-def$ $\lesssim$ 0.4, while it increases
to $HI-def$ $\simeq$ 1 in the core of the Virgo cluster.

Obviously, the use of a complex star formation history characterised by two more free parameters ($QF$ and $QA$) rather than a 
simple secular evolution increases the quality of the fit in all galaxies ($\chi^2_r(T)$/$\chi^2_r(NT)$ $\leq$ 1). Figure
\ref{chiv} shows, however, that the quality of the fit increases significantly in the most HI-deficient galaxies of the sample
($HI-def$ $\gtrsim$ 0.5) and in those objects where the quenching factor is important ($QF$ $>$ 0.8). We also tested whether
$QF$ and $QA$, the two free parameters that we want to analyse in this work, are correlated one another, thus witnessing a possible
strong degeneracy in these outputs of the model. We do not see any clear trend between these two variables. 

\subsection{The quenching factor}

  \begin{figure}
   \centering
   \includegraphics[width=14cm]{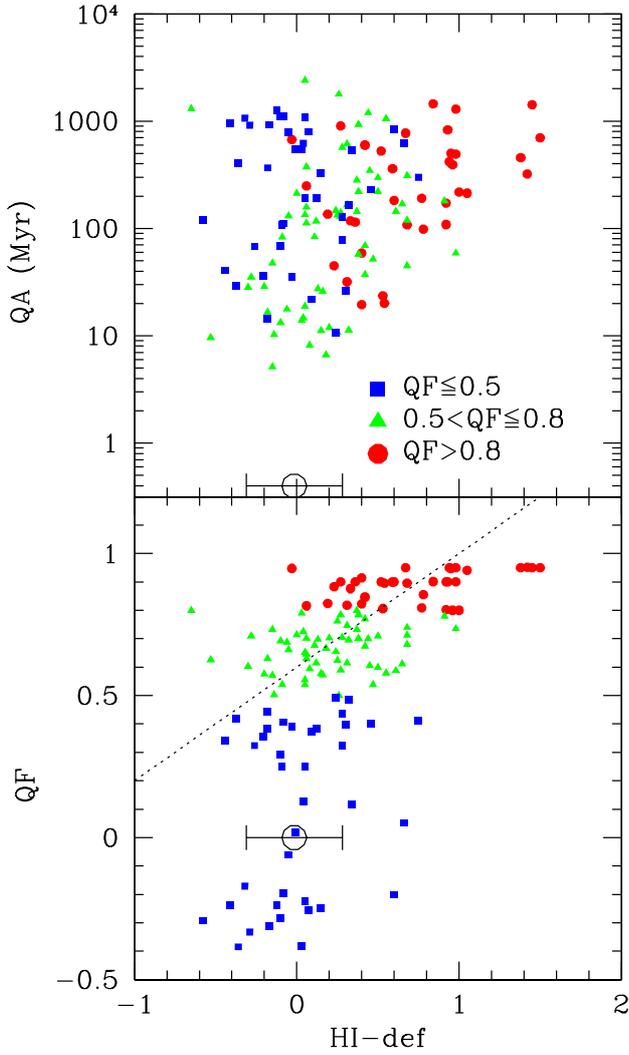}
   \caption{Relationship between the quenching age $QA$ (upper panel) and the quenching factor $QF$ (lower panel) and the
   HI-deficiency parameter for those objects with $\chi_r^2$ $\leq$ 3. Blue filled squares, green triangles, 
   and red circles are for galaxies where the derived quenching factor is $QF$ $\leq$ 0.5, 
   0.5$<$ $QF$ $\leq$ 0.8, and $QF$ $>$ 0.8, respectively. The large black empty symbol shows the mean value and the distribution 
   in HI-deficiency for galaxies with a $QF$ $\leq$ 0.5 (for which the quenching factor and quenching age are arbitrarily
   set to $QF$=0 and $QA$ = 0.4 Myr). The dotted line in the lower panel indicates the best fit found by Ciesla et al. (2016).
    }
   \label{QFA}%
   \end{figure}

Figure \ref{QFA} shows the relationship between the quenching factor $QF$ (lower panel) and the
HI-deficiency parameter. The quenching factor increases with the HI-deficiency parameter, indicating that 
the activity of star formation of the sample galaxies is significantly reduced once the HI gas content is removed 
during the interaction with the hostile cluster environment (Ciesla et al. 2016; the Spearman correlation coefficient derived for the 95 galaxies with $QF$ $>$ 0.5 and $\chi_r^2$ $\leq$ 3 
is $\rho$ = 0.58). The median quenching factor of HI-deficient galaxies ($HI-def$ $>$ 0.4) is
$<QF_{HI-def>0.4}>$ = 0.80 $\pm$ 0.11 for $<HI-def>$ = 0.68 $\pm$ 0.20, roughly indicating that the activity of star formation of
a perturbed galaxy is reduced by $\sim$ 80\% when the gas content drops by a factor of $\sim$ 5. For comparison, 
all the early-type galaxies of the sample with available spectroscopic and UV data have a quenching factor $QF$ =1.
As discussed in Appendix A, these results are robust vs. the adoption of different star formation histories.

\subsection{The quenching age}

Figure \ref{QFA} also shows the relationship between the quenching age $QA$ and the HI-deficiency parameter ($\rho$ = 0.48). There is a clear relationship
between the two variables, in particular if only those galaxies where $QA$ is securely determined ($QF$ $>$ 0.5) are considered. 
The activity of star formation has been reduced $\sim$ 1 Gyr ago in the most HI-deficient objects of the sample ($HI-def$ $\gtrsim$ 0.8). 

   \begin{figure}
   \centering
   \includegraphics[width=14cm]{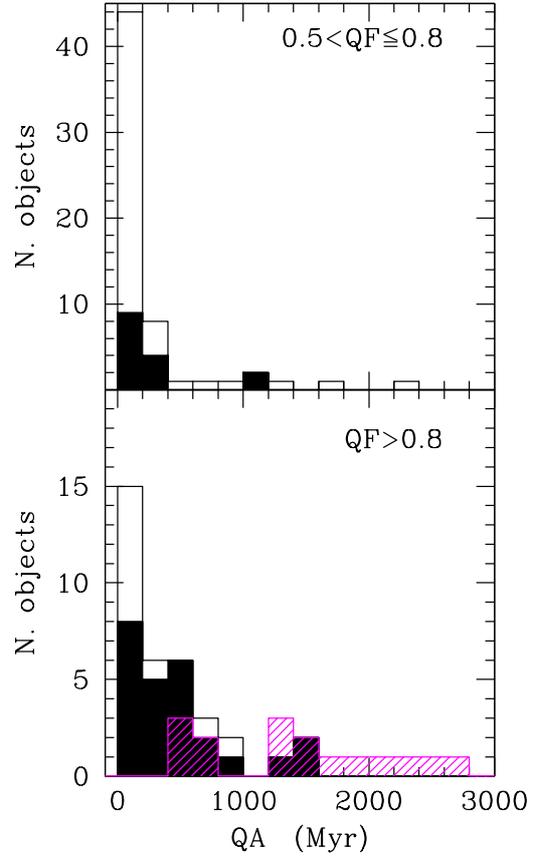}
   \caption{Distribution of the quenching age parameter $QA$ for galaxies with  $\chi_r^2$ $\leq$ 3 and a quenching factor 0.5 $<$ $QF$ $\leq$ 0.8 (upper panel)
   and $QF$ $>$ 0.8 (lower panel). The magenta histogram is for early-type galaxies, the empty histogram for all late-type galaxies, and the black shaded 
   histogram for HI-deficient ($HI-def$ $>$ 0.4) late-type systems.
    }
   \label{QAdist}%
   \end{figure}

Figure \ref{QAdist} shows the distribution of the quenching age parameter $QA$. The median quenching
age of late-type galaxies is $QA$ = 135 Myr for 0.5 $<$ $QF$ $\leq$ 0.8 and $QA$ = 250 Myr for $QF$ $>$ 0.8, while that of 
early-type galaxies is $QA$ = 1320 Myr (see Table A.1). As dicussed in Appendix A, these numbers do not change significantly if the SED 
fitting is done leaving the rotational velocity as a free parameter. They increase by a factor of $\sim$ 2 in the case that
the quenching mechanism is parametrised with a smoothly declining star formation history.

\section{Discussion}

\subsection{Comparison with previous studies}

As discussed in sect. 5.2, the quenching ages determined in this work are comparable to those derived using independent techniques on 
dozen of well studied galaxies in the Virgo cluster. These timescales are also consistent 
with a rapid transformation of gas rich systems into quiescent objects derived from the analysis of the $NUV-i$ colour magnitude relation 
in the Virgo cluster and its surrounding regions (Boselli et al. 2014c; see also Cortese \& Hughes 2009, Hughes \& Cortese 2009) and with the most recent estimates of the galaxy 
infall rate in the cluster ($\sim$ 300 Gyr$^{-1}$ for galaxies with $M_{star}$ $\gtrsim$ 10$^8$ M$_{\odot}$, 
Boselli et al. 2008; Gavazzi et al. 2013a). These short timescales are also consistent with the presence of a
long tail of extraplanar ionised gas associated to NGC 4569 (Boselli et al. 2016), one of the most HI-deficient Virgo cluster 
galaxies ($HI-def$ = 1.05; Boselli et al. 2014a) characterised by a quenched star formation activity ($QF$ = 0.90). The tail of ionised gas
witnesses a stripping process still ongoing.

Direct observational evidences of a rapid quenching of the star formation activity of cluster galaxies are the presence of poststarburst
objects in the periphery of nearby clusters such as Coma and A1367 (Poggianti et al. 2001a,b, 2004; Smith et al. 2008, 2009, 2012; 
Gavazzi et al. 2010). These objects are frequent also in other nearby (Fritz et al. 2014; Vulcani et al. 2015) 
and intermediate to high redshift clusters (Dressler et al. 1999, 2013; Balogh et al. 1999, 2011; 
Poggianti et al. 2004, 2009; Tran et al. 2007; Muzzin et al. 2012, 2014; Mok et al. 2013; Wu et al. 2014). The frequency of
these objects with a recently truncated star formation activity increases with decreasing stellar mass (Boselli \& Gavazzi 2014). 
Evidence for a rapid transformation at intermediate redshift comes also from the analysis of the H$\alpha$ luminosity 
function (e.g. Kodama et al. 2004).

The analysis of large samples of nearby galaxies in different environments, from groups to massive clusters, extracted from complete surveys such
as the SDSS, GALEX, or GAMA, combined with the predictions of cosmological simulations or semi analytical models of galaxy evolution, indicate
that the activity of star formation gradually and moderately decreases once galaxies become satellites of more massive galaxies on 
relatively long timescales (2-7 Gyr) (McGee et al. 2009; von der Linden et al. 2010; De Lucia et al. 2012).
Similar results have been obtained also through the analysis of limited samples of nearby galaxies
with available multifrequency data (Wolf et al. 2009; Haines et al. 2015; Paccagnella et al. 2016).
More recent works suggest that after this mild decrease, the activity rapidly drops on short timescales (0.2-0.8 Gyr) 
once the atomic and molecular gas content is consumed via star formation (Wetzel et al. 2012, 2013; Wijesinghe et al. 2012; Muzzin et al. 2012).  
Pre-processing in infalling groups is thus
probably important in shaping the evolution of galaxies now members of massive clusters (Dressler 2004).
The only indication for a very inefficient quenching of the star formation 
activity ($>$ 9.5 Gyr) of dwarf galaxies ($10^{8.5}$ $<$ $M_{star}$ $<$ $10^{9.5}$ M$_{\odot}$) in relatively dense environments 
comes from the analysis of SDSS data combined with cosmological simulations done by Wheeler et al. (2014).

\subsection{Identification of the perturbing mechanism}

The typical timescales derived in the previous section and in Appendix A can be compared to those predicted by models and simulations 
for the identification of the dominant stripping mechanism in the nearby Virgo cluster. 
We recall that, given the analytical form of the star formation history used in this work (eq. 4), the quenching ages derived in the
previous sections are not the typical timescales necessary to reduce the activity of star formation by a factor $QF$, but just the 
lookback time to the quenching epoch. However, since they are typically $QA$ $\lesssim$ 1.3 Gyr for $QF$ $\leq$ 1, they indicate that
the timescale for quenching the activity of star formation is relatively short. This is also the case when a smoothly declining quenching process
is assumed (see Appendix A). 

The typical timescale for galaxy-galaxy interactions is longer than the age of the universe in dense environments such as the Virgo cluster 
given the large velocity dispersion of this massive system (Boselli \& Gavazzi 2006). Galaxy harassment is more efficient than galaxy-galaxy
interactions given the contribution of the potential well of the cluster. The simulations of Mastropietro et al. (2005), tuned to reproduce 
a Virgo cluster like environment for galaxies of a few 10$^9$ M$_{\odot}$, 
indicate that the typical timescale necessary to heat a disc galaxy and form a pressure supported system is of a few Gyr. Unfortunately
these simulations do not make any prediction on the evolution of the star formation activity, but we can expect that these timescales are
comparable to those required to heat the perturbed galaxy. Again, these timescales are longer than those derived in the previous section 
just because the typical relaxation time in clusters such as Virgo is comparable to the age of the Universe.

\begin{figure*}
   \centering
   \includegraphics[width=14cm]{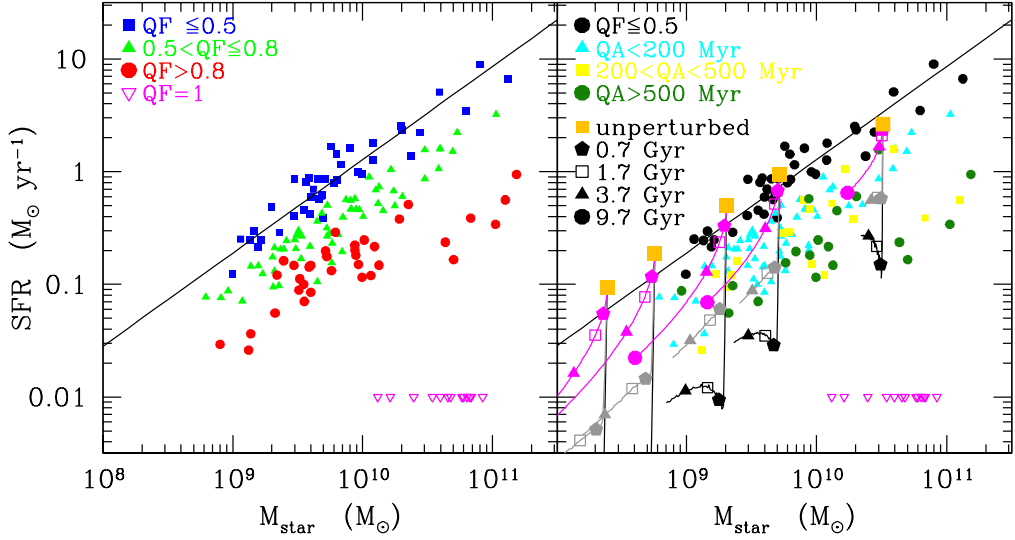}
   \caption{Relationship between the present day star formation rate and the stellar mass (main sequence) derived from the SED fitting code. 
   Left panel: symbols are coded according to the quenching factor of the analysed galaxies. Blue filled squares, green triangles, 
   and red circles are for galaxies where the derived quenching factor is $QF$ $\leq$ 0.5, 
   0.5$<$ $QF$ $\leq$ 0.8, and $QF$ $>$ 0.8, respectively. Magenta triangles indicate upper limits to the star formation rate 
   of early-type galaxies. The solid line shows the bisector fit to the data for unperturbed galaxies ($QF$ $\leq$ 0.5). Right panel: 
   same relation with colour simbols coded according to the quenching age. Black filled dots are for unperturbed galaxies ($QF$ $\leq$ 0.5), cyan triangles, yellow squares, and 
   dark green dots for galaxies with a quenching age $QA$ $<$ 200 Myr, 200 $<$ $QA$ $<$ 500 Myr, and $QA$ $>$ 500 Myr, respectively. 
   The large orange filled squares indicate 
   the models of Boselli et al. (2006, 2014c) for unperturbed galaxies of spin parameter $\lambda$=0.05 and rotational velocity 70, 100, 130, 170, 
   and 220 km s$^{-1}$. The magenta lines indicate the starvation models. The black and grey lines show the ram pressure stripping models for a stripping efficiency 
   $\epsilon_0$ = 1.2 M$_{\odot}$ kpc$^{-2}$ yr$^{-1}$ and $\epsilon_0$ = 0.4 M$_{\odot}$ kpc$^{-2}$ yr$^{-1}$, respectively.
   Different symbols along the models indicate the position of the model galaxies at a given look-back time from the beginning of the interaction. }
   \label{main}%
   \end{figure*}

In a starvation scenario, a galaxy entering a dense environment looses its halo of hot gas stopping the infall of fresh material. 
The activity of star formation gradually decreases after the consumption of the cold atomic and molecular gas located 
on the disc (Larson et al. 1980). The typical timescale for gas consumption of the HRS galaxies is $\simeq$ 4 Gyr (Boselli et al. 2014d), 
and drops to $\gtrsim$ 3.0-3.3 Gyr in gas deficient objects (Boselli et al. 2014b). These timescales, however, should be considered as 
lower limits since the star formation activity of the perturbed galaxies gradually decreases once the gas is removed, thus extending 
the time during which galaxies can still form stars. Tuned models of galaxy starvations of the representative 
HRS galaxy NGC 4569 (Boselli et al. 2006) or of objects with a wide range of stellar masses (Boselli et al. 2014c) consistently indicate
that the starvation process requires very long timescales ($\gtrsim$ 6 Gyr) to significantly reduce the star formation activity
of the perturbed galaxies. Long timescales for starvations to be efficient are also indicated by numerical simulations (e.g. Bekki et al.
2002; McGee et al. 2014). This is depicted in Fig. \ref{main}, where the main sequence diagram derived using the output 
parameters of the SED fitting is compared to the predictions of the evolutionary models of Boselli et al. (2006, 2008, 2014c) for starvation and
ram pressure stripping. Figure \ref{main} clearly shows that, even if the process started $\sim$ 10 Gyr ago, starvation would never be able to 
reduce the star formation activity of the perturbed galaxies by $QF$ $\gtrsim$ 0.5.

\begin{figure}
   \centering
   \includegraphics[width=10cm]{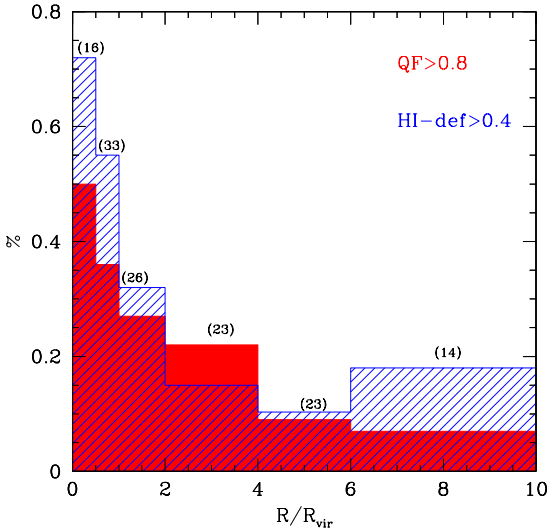}
   \caption{Variation of the fraction of late-type galaxies with a quenching factor $QF$ $>$ 0.8 (red histogram) and an HI-deficiency
   $HI-def$ $>$ 0.4 (blue shaded histogram) and the distance from 
   the centre of the Virgo cluster, in units of virial radii. The number galaxies for each bin of distance is given in parenthesis. 
    }
   \label{fraction}%
   \end{figure}

Much shorter timescales for an efficient quenching of the star formation activity are instead predicted by hydrodynamic
simulations (Roediger \& Henseler 2005; Roediger \& Bruggen 2006, 2007; Tonnesen et al. 2007; Tonnesen \& Bryan 2009; Bekki 2009)
and chemo-spectrophotometric models (Boselli et al. 2006; 2008; 2014c; Cortese et al. 2011) 
of ram pressure stripping in clusters of mass comparable to Virgo. The predictions of these chemo-spectrophotometric models 
derived for two different ram pressure stripping efficiencies, $\epsilon_0$ = 1.2 M$_{\odot}$ kpc$^{-2}$ yr$^{-1}$ necessary to
reproduce the observed truncated profiles of the Virgo cluster galaxy NGC 4569 (Boselli et al. 2006), 
and $\epsilon_0$ = 0.4 M$_{\odot}$ kpc$^{-2}$ yr$^{-1}$, the mean stripping efficiency for galaxies with the typical velocity 
dispersion within the Virgo cluster (Boselli et al. 2008), are compared to the data in Fig. \ref{main}. The observed decrease 
in the star formation activity of HI-deficient galaxies (Boselli et al. 2015) in the main sequence relation can be explained by 
by ram pressure stripping, a result similar to the one derived from the analysis of the UV-to-optical
colour magnitude relation (Cortese \& Hughes 2009; Hughes \& Cortese 2009; Boselli et al. 2014c,d). 
The large number of objects below the main sequence drawn by unperturbed systems, often taken as an indication that the 
perturbing process lasts several Gyrs, is due to the large infall rate of star forming systems observed in Virgo ($\simeq$ 300 Gyr$^{-1}$; Tully \& Shaya 1984; 
Boselli et al. 2008; Gavazzi et al. 2013a). A further evidence in agreement with the ram pressure
stripping scenario is given in Fig. \ref{fraction}, where the fraction of late-type galaxies with a quenching factor $QF$ $>$ 0.8 
and an HI-deficiency $HI-def$ $>$ 0.4 are plotted versus the distance from the core of the Virgo cluster. Figure \ref{fraction}
clearly shows that both fractions drop by a factor of $\simeq$ 5 from the core of the cluster ($R/R_{vir}$ $<$ 0.5), where the hot X-ray gas
is detected by \textit{ROSAT} (Bohringer et al. 1994), to the cluster periphery ($R/R_{vir}$ $\gtrsim$ 4), in agreements with models and
simulations (Tonnesen et al. 2007; Bahe et al. 2013; Cen et al. 2014). 
The recent observations of nearby clusters with 4-8 m class telescopes equipped with wide field panoramic detectors and 
narrow band filters is revealing the presence up to the cluster virial radius of several star forming galaxies with long tails of 
ionised gas typically produced by a ram pressure stripping event (Gavazzi et al. 2001; Sun et al. 2007; Yagi et al. 2010; 
Fossati et al. 2012). Two galaxies with long tails of
HI gas have been also observed in the periphery of A1367, outside the virial radius (Scott et al. 2012). The radial variation of the 
fraction of quenched and HI-deficient galaxies is consistent with the one observed in the Virgo cluster out to large distances by Gavazzi et al. (2013a) and 
Boselli et al. (2014c), a further evidence that quenching is a rapid process (Boselli et al. 2014c).

Short timescales are also consistent with a recent formation of the faint end of the red sequence
(De Lucia et al. 2007, 2009; Stott et al. 2007, 2009; Gilbank \& Balogh 2008). 

  \begin{figure}
   \centering
   \includegraphics[width=14cm]{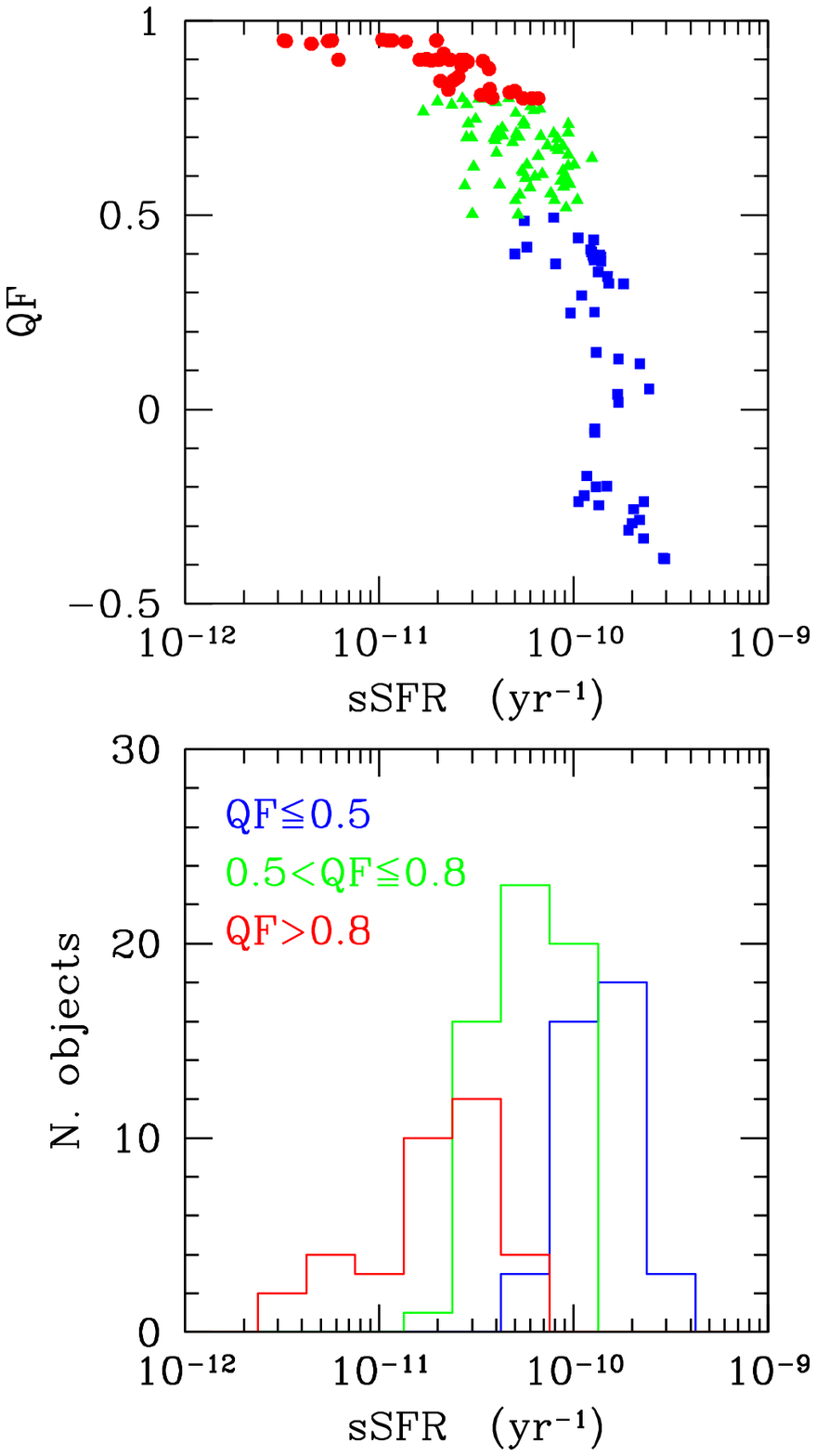}
   \caption{Upper panel: relationship between the quenching factor $QF$ and the specific star formation activity $sSFR$
   for the late type galaxies of the sample with a reduced $\chi_r^2$ $\leq$ 3. Blue filled squares, green triangles, 
   and red circles are for galaxies where the derived quenching factor is $QF$ $\leq$ 0.5, 
   0.5$<$ $QF$ $\leq$ 0.8, and $QF$ $>$ 0.8, respectively. Lower panel: $sSFR$ distribution for different quenching factors.
    }
   \label{SSFR}%
   \end{figure}

\subsection{Rapid or slow environmental quenching?}

We often read in the literature that the environmental quenching acts on long timescales 
(McGee et al. 2009; von der Linden et al. 2010; De Lucia et al. 2012; Taranu et al. 2014; Haines et al. 2015; Paccagnella et al. 2016). This apparent  
disagreement with our results comes from the fact that different definitions of quenching timescales 
are often used. In those studies based on the combined analysis of statistical samples extracted from 
all sky surveys such as the SDSS with the predictions of cosmological simulations or semi analytical models, 
the quenching timescale is often defined as the time necessary to stop the activity of star formation of a galaxy 
once it becomes satellite of a larger system. Since the recent accretion of late-type galaxies in nearby clusters 
occurs principally through smaller groups (Gnedin 2003; McGee et al. 2009; De Lucia et al. 2010), this timescale is necessary long
even if the quenching of the activity happens on short timescales only once the galaxy fall in the main cluster structure. Indeed, the 
analysis of Wetzel et al. (2013) suggests a different evolution of the star formation activity of galaxies, first mildly declining
on long timescales (2-4 Gyr), then with a rapid decrease (0.2-0.8 Gyr) ('delayed-then-rapid'). 
The timescales that we estimate in this work, as those derived for other Virgo cluster galaxies
as described in sect. 5.2, are rather representative of the timescale since the first stripping of the cold gas on the disc,
which generally occurs once the galaxy is close to the pericenter along its orbit within the cluster. 
Recently, Oman \& Hudson (2016) studied the quenching timescale and efficiency in massive clusters ($M_{cluster}$ $>$ 10$^{13}$ M$_{\odot}$) 
by comparing N-body simulations with the observed properties of SDSS galaxies.
This work indicates that the quenching is 100\% ~efficient ($QF$ = 1) and occurs in the core of the cluster, at or near the first pericenter approach. 
These results convincingly point at ram pressure stripping as the main quenching mechanism in massive haloes. 
While their quenching timescale is the typical value for a population of galaxies in clusters, their results 
are in excellent agreement with our study of the histories of indivudual galaxies in Virgo.
 
It is also worth remembering that the dominant perturbing mechanism is expected to change as a function of the mass of the 
infalling galaxy and on quantities that strongly depend on the total dynamical mass of the cluster such as the velocity dispersion of galaxies within the cluster, 
and the density and temperature of the intracluster gas. These quantities might also vary locally within the cluster 
since the distribution of the intracluster medium is not uniform in terms of density and turbulence. Since all these parameters 
change also with redshift, we expect to have different results for different environments at different epochs,
as indeed observed (e.g. Balogh et al. 2016). 

A further possible origin of the apparent disagreement between the short timescales derived in this work with 
those given in the literature might be due to the 
definition of quenched galaxies. Here we consider quenched galaxies those objects with an activity of star formation reduced by 
a factor of $QF$ $>$ 0.5 or $QF$ $>$ 0.8. These are still star forming late-type galaxies. As depicted in Fig. \ref{SSFR}, 
their typical specific star formation rate, defined as the current star formation rate devided by the total stellar mass given by the fit, 
ranges from a few 10$^{-12}$ yr$^{-1}$ to $sSFR$ $\lesssim$ 10$^{-10}$ yr$^{-1}$.
A typical example is the massive NGC 4569 in the core of the Virgo cluster ($sSFR$ = 4.5 $\times$ 10$^{-11}$ yr$^{-1}$), charaterised by a prominent H$\alpha$ 
emission witnessing an ongoing star formation activity. In SDSS based works such as Wetzel et al. (2012, 2013), 
quenched galaxies are those with a specific star formation rate $sSFR$ $\lesssim$ 10$^{-11}$ yr$^{-1}$, a much more stringent 
definition than the one used in this work. This difference is important in the framework of galaxy evolution. 
In a ram pressure stripping scenario, models and simulations indicate that most of the gas is stripped from the disc of spiral 
galaxies on short timescales ($\lesssim$ 0.2 Gyr) when the galaxy is at the pericenter along its orbit (Roediger \& Henseler 2005; 
Roediger \& Bruggen 2006, 2007; Tonnesen et al. 2007; Tonnesen \& Bryan 2009). At this stage, however, there is still gas to 
sustain a moderate star formation activity on the inner disc of the perturbed galaxies (Boselli et al. 2006; Tonnesen \& Bryan 2012;
Bekki 2014; Roediger et al. 2015). Their quenching factor is thus $QF$ $<$ 1 and their specific star formation
activity $sSFR$ $\gtrsim$ 10$^{-11}$ yr$^{-1}$. The complete stripping of the gas, thus the total quenching of the star formation activity 
($QF$ = 1), requires longer timescales (1-3 Gyr), consistent with what found for the 22 early-type galaxies of the HRS.
These timescales are fairly comparable to the timescale for the rapid quenching of the star formation 
activity of galaxies in rich environments found by Wetzel et al. (2012, 2013) and Oman \& Hudson (2016)
in the local universe and by Muzzin et al. (2014) and Mok et al. (2014) at $z$ $\sim$ 1.

It seems thus that the origin of most of the tension between these works is in the definition of the quenching time, here defined as the 
epoch when galaxies drastically reduced ($\gtrsim$ 50 \%) their activity of star formation, in Wetzel et al. (2012, 2013) and in other works
the time since a galaxy $\it{first}$ became a satellite of a massive halo.

We conclude by mentioning that this analysis indicates that the quenching time
can be accurately derived using our SED fitting technique only in 30\%~ of the sample (those galaxies with $QF$ $>$ 0.8),
a value comparable to the number of HI-deficient objects ($HI-def$ $>$ 0.4; 33\%) or of late-type galaxies located 
within one virial radius of Virgo (36\%). Whenever the perturbing mechanism is milder than in
Virgo, as it is probably the case in less massive haloes,
we expect a less efficient quenching episode ($QF$ $<<$ 1). For these galaxies it is hard to break all the degeneracies
between star formation history, IMF, metallicity, and  
dust attenuation, and thus derive accurate quenching timescales even using a UV-to-FIR SED fitting code as done in this work.

\section{Conclusion}
 
We study the star formation history of the \textit{Herschel} Reference Survey, a $K$-band-selected, volume-limited complete sample of 
nearby galaxies using the CIGALE SED fitting code. The sample includes objects in the Virgo cluster and is thus
perfectly suited for understanding the role of the environment on galaxy evolution. To quantify the perturbation
induced by the interaction with the hostile environment on the star formation activity of cluster galaxies, we
adopt a truncated star formation history where the secular evolution is parametrised using the chemo-spectrophotometric 
physically justified models of Boissier \& Prantzos (2000). 
To constrain any possible abrupt variation of the star formation activity of the Virgo cluster galaxies, 
we combine the UV to far infrared photometric data (20 bands) with age sensitive Balmer absorption line indices extracted from 
medium resolution ($R$ $\sim$ 1000) integrated spectroscopy (3 bands) and H$\alpha$ imaging data (1 band). H$\alpha$ fluxes are 
compulsory to constraint variations occurred at very recent epochs ($\lesssim$ 100 Myr). The best fit to the data gives
the quenching factor $QF$ and the quenching age $QA$. The first parameter quantifies how much the star formation activity has been
reduced during the interaction ($QF$=0 for unperturbed systems, $QF$ = 1 when the star formation activity has been completely stopped), 
while $QA$ gives the lookback time of the epoch of the quenching episode. We checked the reliability of the output parameters 
using mock catalogues of simulated galaxies and different star formation histories.

The analysis of the sample brought to the following results:\\
1) The quality of the SED fitting significantly increases using a truncated star formation history in all the
HI-deficient galaxies of the sample, where the interaction with the hostile environment removed a large fraction of the 
cold gas necessary to feed star formation.\\
2) In these HI-deficient objects the activity of star formation is reduced by a factor of 50\% $\leq$ $QF$ $<$ 80\% 
on timescales of $\simeq$ 135 Myr, and $QF$ $\geq$ 80\% on timescales of $\simeq$ 250 Myr, while it is fully stopped 
after $\simeq$ 1.3 Gyr in early-type galaxies. These timescales are a factor of $\simeq$ 2 longer if a smoothly declining quenching process 
is assumed.\\
3) The fraction of quenched late-type galaxies ($QF$ $\geq$ 80\%), as the fraction of HI-deficient objects, 
decreases by a factor of $\simeq$ 5 from the core of the Virgo cluster ($R$/$R_{vir}$ $\leq$ 0.5) to the cluster periphery 
($R$/$R_{vir}$ $\geq$ 4).

The disagreement with the results obtained from the analysis of large samples such as the SDSS (e.g. Wetzel et al. 2012, 2013), which indicate 
longer timescales with a delayed-then-rapid quenching process, is only apparent since 
our timescales are representative of the rapid decrease of the star formation activity occured when galaxies enter the rich Virgo cluster, while those 
give in the literature of the time since a galaxy became a satellite of a more massive halo. Furthermore, our are the timescales necessary to
reduce the star formation activity by a factor of $\gtrsim$ 50-80\%, while those given in the literature to fully stop the activity
of the perturbed galaxies. 
All these results are consistent with a rapid quenching of the star formation activity of the late-type galaxies
recently accreted on the Virgo cluster as predicted by ram pressure stripping models. These results discard inefficient mechanisms such as
starvation, which require very long ($\gtrsim$ 6 Gyr) timescales to significantly quench the star formation activity of the perturbed
galaxies.

\begin{acknowledgements}
We thank the anonymous referee for constructive comments which improved the clarity of the manuscript.
This research has been financed by the French national program PNCG.
M. Fossati acknowledges the support of the Deutsche Forschungsgemeinschaft via Project ID 3871/1-1. 
L. Ciesla acknowledges funding from the European Union Seventh Framework Programme (FP7/2007-2013) under grant agreement n 312725.
This research has made use of the NASA/IPAC Extragalactic Database (NED) 
which is operated by the Jet Propulsion Laboratory, California Institute of 
Technology, under contract with the National Aeronautics and Space Administration.

\end{acknowledgements}

\begin{appendix}

\section{Exploring different star formation histories}

\begin{figure}
   \centering
   \includegraphics[width=9cm]{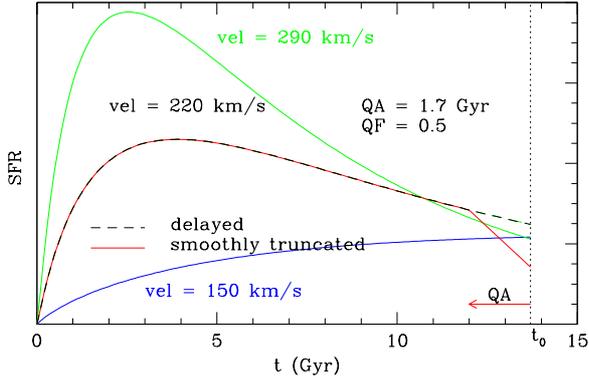}
   \caption{The parametric star formation history for a galaxy with a rotational velocity of 220 km s$^{-1}$ (black dashed line),
   290 km s$^{-1}$ (green solid line) and 150 km s$^{-1}$ (blue solid line).
   The black dashed line shows the delayed star formation history given in eq. (1), the red solid line the smoothly declining one (eq. A.1)
   for a galaxy of rotational velocity 220 km s$^{-1}$. This figure can be compared to
   Fig. \ref{SFH}.
   }
   \label{SFHsm}%
   \end{figure}

To test the reliability of the output parameters of the SED fitting code we have also derived the quenching ages and the quenching factors using two other star formation
histories. In the first one, we use the same truncated star formation history given in eq. 4 but we leave the rotational velocity of the galaxy as a free parameter
and derive $QF(vel)$ and $QA(vel)$. In the second we 
fix the rotational velocity of the galaxies but we use a smoothly declining star formation history where the decrease of the activity is constant from the beginning of the
interaction to the present epoch, as depicted in Fig. \ref{SFHsm} ($QF(SM)$ and $QA(SM)$):

	\begin{equation}
	$$
	SFR(t) = \left\{ \begin{array}{ll}
	SFR(t)_{secular} &\mbox{ if $t_0$-$t$ $\geq$ QA} \\
	SFR(t_0-QA)_{secular} \\
	\times [1-QF+\frac{QF}{QA}\times (t_0-t)] &\mbox{ if $t_0$-$t$ $<$ QA} 
	\end{array} \right.
	$$
	\end{equation}

\begin{figure*}
   \centering
   \includegraphics[width=14cm]{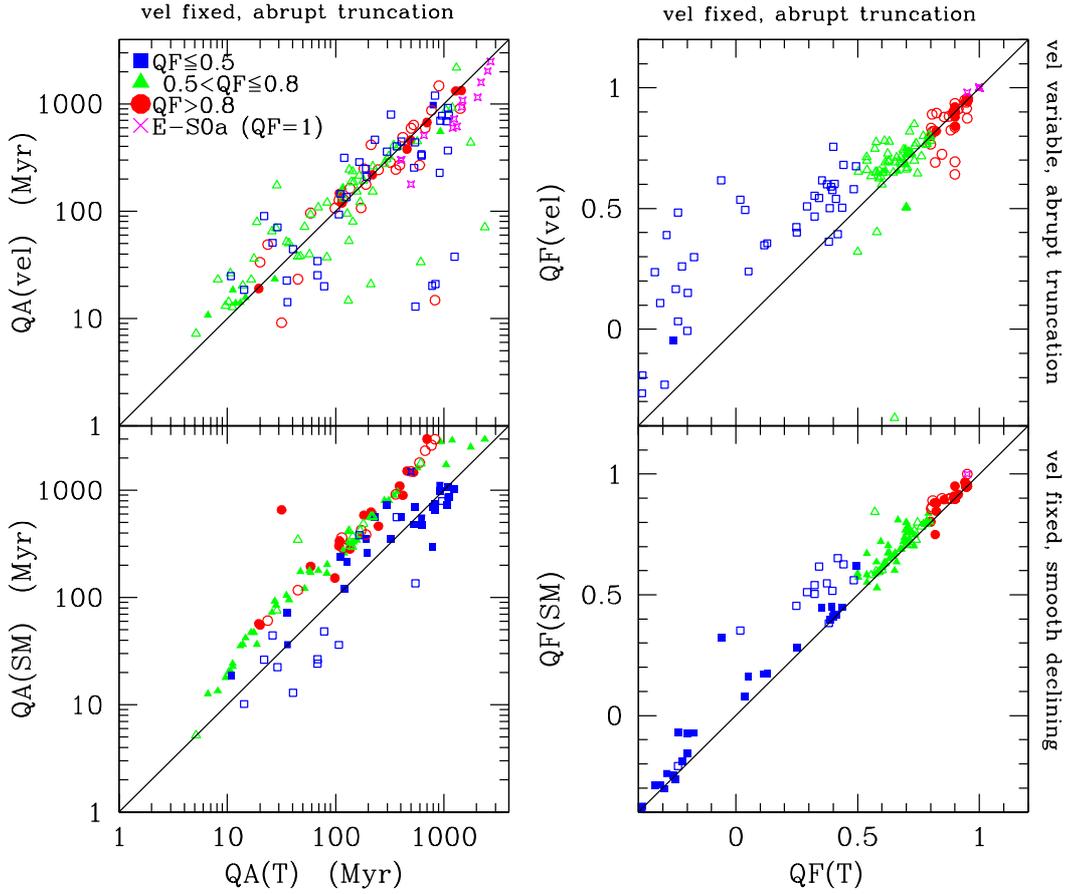}
   \caption{Left panel: relationship between the quenching age parameter derived using a truncated star formation history (eq. 4) with the rotational velocity as
   a free parameter ($QA(vel)$, upper panel) or using a smoothly declining star formation history (eq. A.1) with fixed rotational velocity ($QA(SM)$, lower panel) vs.
   the one derived using a truncated star formation history (eq. 4) determined using a fixed value for the rotational velocity $QA(T)$. Filled symbols and crosses indicate
   galaxies where the reduced $\chi^2_r$ in the truncated model with fixed velocity is smaller than the one derived with the smoothly declining star formation history or the 
   truncated one with the rotational velocity as a free parameter, empty symbols for $\chi^2_r(T)/\chi^2_r(SM,vel)$ $>$ 1. Blue squares, green triangles, 
   and red circles are for late-type galaxies with a quenching factor $QF$ $\leq$ 0.5, 
   0.5$<$ $QF$ $\leq$ 0.8, and $QF$ $>$ 0.8, respectively, while the magenta crosses and empty stars for early-type galaxies. The black solid line shows the 1:1 relation.
   Only galaxies with a $\chi_r^2$ $\leq$ 3 are plotted.
   Right panel: same relationships for the quenching factor.
   }
   \label{smooth}%
   \end{figure*}

Figure \ref{smooth} shows the relationship between these newly derived parameters and the quenching ages and quenching factors derived in the paper assuming a fixed rotational
velocity and the truncated star formation history given in equation 4.

Figure \ref{smooth} shows that the quenching factors derived using these three different assumptions on the star formation history and on the rotational velocity of the galaxies
are very consistent provided that $QF$ $\gtrsim$ 0.5, i.e. in the range of parameters analysed in this work. Figure \ref{smooth} also shows that the 
quenching ages derived using a truncated star formation history (eq. 4) are very consistent regardless the use of a fixed or a variable rotational velocity. 
This is even clearer whenever the 
quenching factor is large (the dispersion decreases going from the blue, to the green, the red, and the magenta symbols). On the contrary, there is a systematic shift in the
quenching ages when these parameters are derived assuming a truncated or a smoothly declining star formation history, the latter giving systematically larger values 
(the median values for the $QA(SM)/QA(T)$ are 2.58, 2.80, and 2.98 for the late-type galaxies with 0.5 $<$ $QF(T)$ $\leq$ 0.8, $QF(T)$ $>$ 0.8, and for the early-types, respectively).
This is expected since the quenching ages derived using a smoothly decling star formation history should be considered as upper limits if the decrease of the activity is more rapid than
the one predicted by eq. A.1.

   \begin{figure*}
   \centering
   \includegraphics[width=14cm]{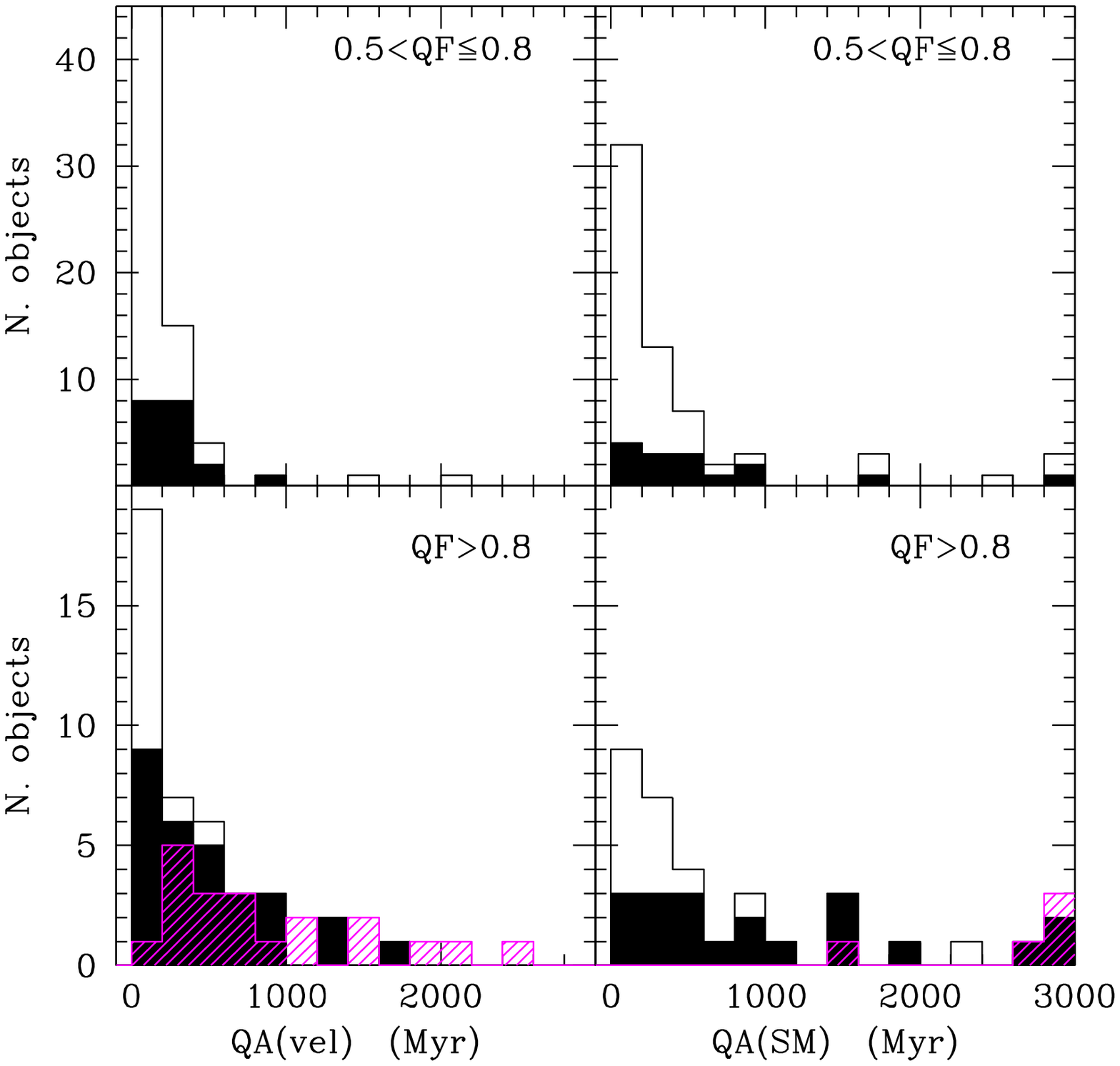}
   \caption{Left panels: distribution of the quenching age parameter $QA(vel)$ for galaxies with  $\chi_r^2$ $\leq$ 3 and a quenching factor 0.5 $<$ $QF(vel)$ $\leq$ 0.8 (upper panel)
   and $QF(vel)$ $>$ 0.8 (lower panel) derived using a truncated star formation history (eq. 4) with the rotational velocity as a free parameter. 
   The magenta histogram is for early-type galaxies, the empty histogram for all late-type galaxies, and the black shaded 
   histogram for HI-deficient ($HI-def$ $>$ 0.4) late-type systems. Right panels: the same distributions derived using a smoothly declining star formation history (eq.
   A.1) and a fixed rotational velocity.
    }
   \label{QAdistvelvar}%
   \end{figure*}

\begin{figure}
   \centering
   \includegraphics[width=12cm]{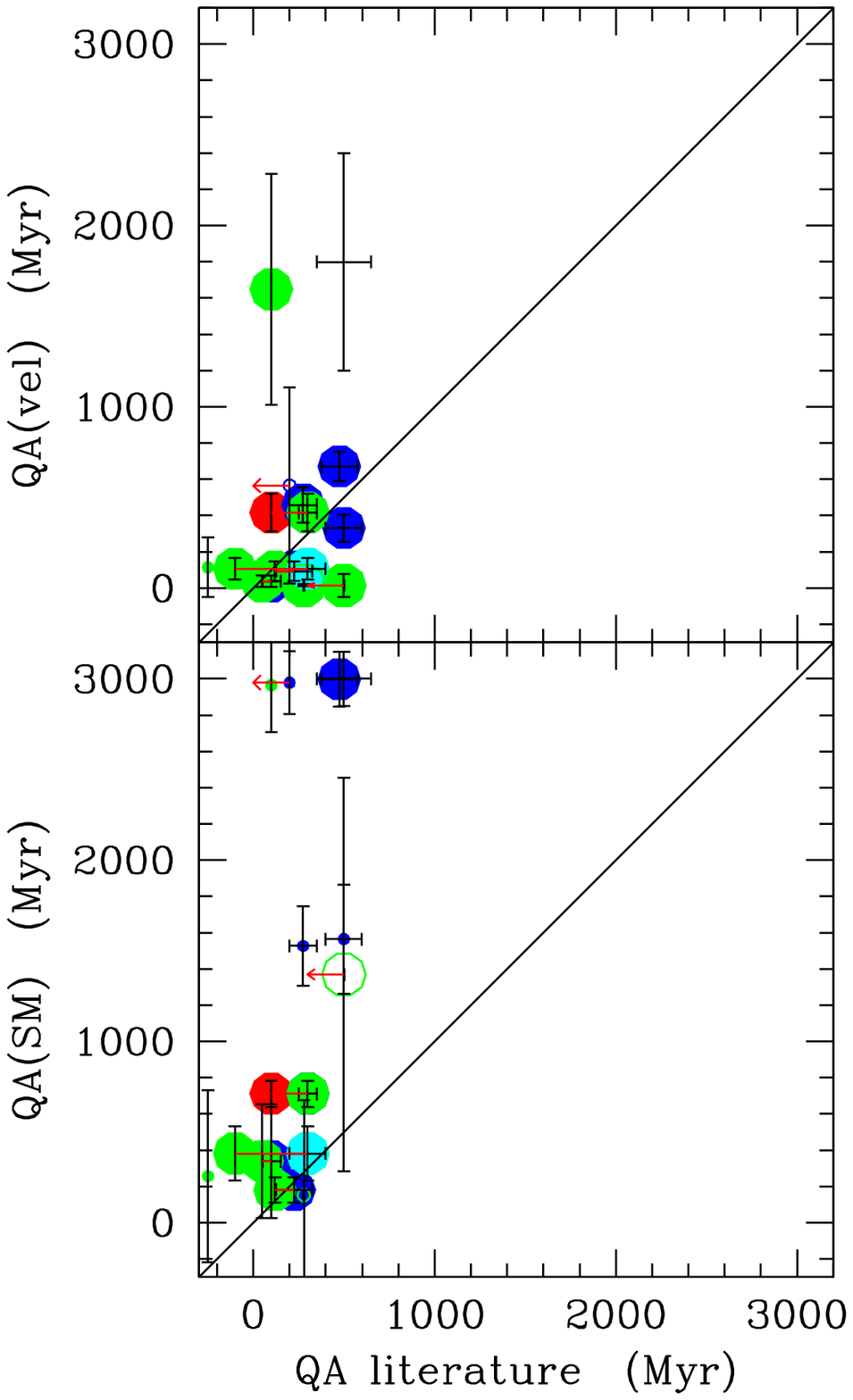}
   \caption{Relationship between the quenching age derived using our SED fitting technique and that derived using multizone 
   chemo-spectrophotometric models of galaxy evolution for NGC 4569 (red dot), dynamical models based on the HI and CO gas kinematics (green dots),
   IFU spectroscopy of the outer disc of some cluster galaxies (blue dots), and photometric data compared to population synthesis models (cyan dots) (see Fig. \ref{known}
   for comparison). 
   In the upper panel the model SEDs are derived using a truncated star formation history (eq. 4) leaving the rotational velocity as a free parameter, 
   in the lower panel the rotational velocity is fixed and the star formation history is smoothly declining (eq. A.1). 
   Filled symbols are used for galaxies with a quenching
   factor $QF$ $>$ 0.5, empty dots for those with $QF$ $\leq$ 0.5. Large symbols are for galaxies with a reduced $\chi_r^2$ $\leq$ 3, 
   small symbols for  $\chi_r^2$ $>$ 3. The red lines connecting two points are used to indicate those galaxies with two independent 
   quenching ages available in the literature. The red arrows indicate upper limits to the quenching age given in the literature. The solid diagonal line indicates the 1:1 relationship. 
    }
   \label{knownsmooth}%
   \end{figure}

Figure \ref{QAdistvelvar} shows the distribution of the quenching ages derived using a truncated star formation history with the rotational velocity as a free parameter
and a smoothly declining star formation history. It can be compared to Fig. \ref{QAdist} derived using a truncated star formation history with fixed rotational velocity.
The median values derived for the three different prescriptions are given in Table \ref{TabQA}. The quenching ages derived using a truncated star formation history
are very similar regardless the use of a fixed or a variable rotational velocity, with the exception of the early-type sample where the $QA$ is $\simeq$ a factor of 2 longer
in the former case. The quenching ages derived using a smoothly declining star formation history are, as expected, a factor of $\simeq$ 2 longer than those derived using a
truncated star formation activity. The significantly larger number of filled symbols with respect to that of the empty ones in the lower panels of Fig. \ref{smooth},
indicating those galaxies where the ratio of the reduced $\chi^2_r$ derived from the SED fitting using a truncated to smoothly declining star formation history is smaller than
unity
($\chi^2_r(T)$/$\chi^2_r(SM)$ $<$1), suggests that the quality of the fit is higher when the truncated approximation is used (the number of free parameters in the two
star formation histories are the same).
This is also evident in the comparison of Fig. \ref{QAdistvelvar} with Fig. \ref{QAdist} 
where only galaxies with a reduced $\chi^2_r$ $\leq$ 3 are plotted. The number of galaxies with a good quality fit is larger when a truncated vs. 
a smoothly declining star formation history is used. A further indication that a truncated star formation history is better suited 
to trace the evolution of the perturbed Virgo cluster galaxies than the smoothly declining one comes from the comparison of Fig. \ref{known} and Fig. \ref{knownsmooth}.
The quenching ages derived using a truncated star formation history with a fixed rotational velocity are those which better reproduce the 
estimates available in the literature and derived using independent techniques for a dozen of galaxies in common (see sec. 5). Although this small subsample of galaxies 
is probably biased since composed of systems expressely selected to have some evidence of an undergoing ram pressure stripping event, we decided to adopt a truncated star formation history and a fixed rotational velocity 
as a reference case for the analysis done in this work. We recall, however, that the quenching ages given by the smoothly declining approximation, which in some cases 
gives better fits, can be taken as upper limits to the effective time taken by these galaxies to reduce their activity by a factor $QF$. The timescales 
to reduce the activity of late-type galaxies by a factor of $QF$ $<$ 1 are $\lesssim$ 500 Myr, and $\lesssim$ 3 Gyr ($\simeq$ 2 crossing times of the Virgo cluster,
Boselli \& Gavazzi 2006) to stop the activity and transform late-type galaxies in lenticulars.

\begin{table}
\caption{Median quenching ages derived assuming different star formation histories for galaxies with $\chi^2_r$ $\leq$ 3. }
\label{TabQA}
{
\[
\begin{tabular}{cccc}
\hline
\noalign{\smallskip}
\hline
Condition				& QA(T)	&  QA(vel)	& QA(SM)        \\
					& Myr	& Myr		& Myr		\\
\hline
Late-types, 0.5$<$ $QF$ $\leq$0.8	& 136	& 127		& 302 \\
Late-types, $QF$ $>$ 0.8		& 248	& 244 		& 423 \\
Early-types ($QF$=1)			& 1319	& 614		& 2997\\
\noalign{\smallskip}
\hline
\end{tabular}
\]
}
\end{table}

\end{appendix}

\end{document}